\documentclass[10pt,twocolumn,twoside,final]{IEEEtran}
\usepackage{graphicx}
\usepackage{float}

\usepackage{fixltx2e}
\usepackage{tabularx}
\usepackage{verbatim}
\usepackage{color}
\usepackage{hyperref}
\usepackage{xmpmulti}
\usepackage{transparent}
\usepackage{fancyhdr}

\usepackage{amssymb,amsmath,graphicx,float}
\usepackage{algorithm, algpseudocode}
\usepackage{booktabs}
\usepackage{caption}
\usepackage{subcaption}
\usepackage{amsthm}
\usepackage{hyperref}
\usepackage{wrapfig}
\usepackage{setspace}
\usepackage{graphicx}
\usepackage{epsfig}
\usepackage[all]{xy}
\usepackage{verbatim}
\usepackage{float}
\usepackage{mathrsfs}  
\usepackage{bm}
\usepackage{multirow}
\usepackage{color,soul}
\usepackage{amssymb,amsmath,graphicx}
\usepackage{amsthm}

\usepackage{array}
\newtheorem{problem}{Problem}

\newtheorem{remark}{Remark}

\newcommand{\norm}[1]{\left\lVert#1\right\rVert}
\newcommand{\tabincell}[2]{\begin{tabular}{@{}#1@{}}#2\end{tabular}}
\let\oldIEEEkeywords\IEEEkeywords
\def\IEEEkeywords{\oldIEEEkeywords\normalfont\bfseries\ignorespaces}
\begin{document}
	
\title{Control Strategies for COVID-19 Epidemic with Vaccination, Shield Immunity and Quarantine: A Metric Temporal Logic Approach}

% \pagestyle{fancy}
% \fancyhf{}
% \rhead{Working paper}
% \rfoot{\thepage}

 	\author{Zhe~Xu\thanks{Zhe~Xu and Bo~Wu are with the Oden Institute for Computational Engineering and Sciences, University of Texas,
 			Austin, Austin, TX 78712, Ufuk Topcu is with the Department
 			of Aerospace Engineering and Engineering Mechanics, and the Oden Institute
 			for Computational Engineering and Sciences, University of Texas,
 			Austin, Austin, TX 78712, e-mail: zhexu@utexas.edu,  bowu86@gmail.com, utopcu@utexas.edu.}, ~Bo~Wu~and Ufuk Topcu}
%  \author{
% 	Zhe Xu, Ufuk Topcu, \\
%   University of Texas at Austin, Austin, TX\\
% 	\{zhexu, utopcu\}@utexas.edu
% }  
\date{\vspace{-5ex}}

\maketitle 

\begin{abstract} 
Ever since the outbreak of the COVID-19 epidemic, various public health control strategies have been proposed and tested against the coronavirus SARS-CoV-2. We study three specific COVID-19 epidemic control models: the susceptible, exposed, infectious, recovered (SEIR) model with vaccination control; the SEIR model with \textit{shield immunity} control; and the susceptible, un-quarantined infected, quarantined infected, confirmed infected (SUQC) model with quarantine control. We express the control requirement in \textit{metric temporal logic} (MTL) formulas (a type of formal specification languages) which can specify the expected control outcomes such as 
``\textit{the deaths from the infection should never exceed one thousand per day within the next three months}'' or ``\textit{the population immune from the disease should eventually exceed 200 thousand within the next 100 to 120 days}''. We then develop methods for synthesizing control strategies with MTL specifications. To the best of our knowledge, this is the first paper to systematically synthesize control strategies based on the COVID-19 epidemic models with formal specifications. We provide simulation results in three different case studies: vaccination control for the COVID-19 epidemic with model parameters estimated from data in Lombardy, Italy; shield immunity control for the COVID-19 epidemic with model parameters estimated from data in Lombardy, Italy; and quarantine control for the COVID-19 epidemic with model parameters estimated from data in Wuhan, China. The results show that the proposed synthesis approach can generate control inputs such that the time-varying numbers of individuals in each category (e.g., infectious, immune) satisfy the MTL specifications. The results also show that early intervention is essential in mitigating the spread of COVID-19, and more control effort is needed for more \textit{stringent} MTL specifications. For example, based on the model in Lombardy, Italy, achieving less than 100 deaths per day and 10000 total deaths within 100 days requires 441.7$\%$ more vaccination control effort than achieving less than 1000 deaths per day and 50000 total deaths within 100 days.
\end{abstract}  

\begin{IEEEkeywords}
	COVID-19 epidemic, vaccination, shield immunity, quarantine, metric temporal logic
\end{IEEEkeywords}

\section{Introduction}
The COVID-19 pandemic \cite{Fauci2020} has caused over 20 million confirmed cases and over 0.74 million deaths globally as of August 12, 2020. Ever since the outbreak of COVID-19, various public health control strategies have been proposed and tested against the coronavirus SARS-CoV-2 \cite{Stewart_magazine_2020}. 

Currently, over 90 vaccines are being developed against SARS-CoV-2 by research teams across the world \cite{vaccines_nature2020}. Besides vaccination, other strategies have also been proposed to control the spread of SARS-CoV-2. In \cite{Weitz_nature}, the authors proposed \textit{shield immunity} to protect the susceptible people from getting infected with SARS-CoV-2. Specifically, shield immunity works by first identifying and deploying recovered individuals who have protective antibodies to SARS-CoV-2, and then increasing the proportion of interactions with
recovered individuals as opposed to other individuals. In \cite{Zhao2020}, the authors analyzed how quarantine has mitigated the spread of SARS-CoV-2 based on a model that differentiates quarantined infected individuals and un-quarantined infected individuals.

Despite the fact that various promising control strategies have been proposed against SARS-CoV-2, such control strategies still suffer from several limitations. (a) The control strategies against SARS-CoV-2 often treat the control inputs (e.g., the shield strength in shield immunity and the quarantine rate in quarantine control) as parameters that stay constant during one stage of time, while in reality such parameters may change on a daily basis with more fine-tuned control. (b) The control inputs in the literature are often tuned manually through trial-and-error instead of being synthesized systematically. (c) There is a lack of specific and formal specifications for the expected effects and outcomes of the control strategies. 
 
To address these limitations, we propose a systematic control synthesis approach for three control strategies against SARS-CoV-2. We use \textit{metric temporal logic} (MTL) formulas to specify the expected control outcomes such as ``\textit{the deaths from the infection should never exceed one thousand per day within the next three months}'' or ``\textit{the population immune from the disease should eventually exceed 200 thousand within the next 100 to 120 days}''. Such temporal logic formulas have been used as high-level knowledge or specifications in many applications in artificial intelligence \cite{zhe_ijcai2019}, robotic control \cite{Verginis2019Icra}, power systems \cite{zhe_control}, etc. 

The proposed control synthesis approach is based on three specific COVID-19 epidemic mitigation models: the susceptible, exposed, infectious, recovered (SEIR) model with vaccination control; the SEIR model with shield immunity control; and the susceptible, un-quarantined infected, quarantined infected, confirmed infected (SUQC) model with quarantine control. We develop methods for synthesizing control strategies based on the three specific COVID-19 epidemic models with MTL specifications. Specifically, we convert the synthesis problem into mixed-integer bi-linear programming or mixed-integer fractional constrained programming problems, and solve the optimization problems using highly efficient solvers \cite{beal2018gekko}.

We provide simulation results in three different case studies: vaccination control for COVID-19 epidemic with model parameters estimated from data in Lombardy, Italy; shield immunity control for COVID-19 epidemic with model parameters estimated from data in Lombardy, Italy; and quarantine control for COVID-19 epidemic with model parameters estimated from data in Wuhan, China. The proposed synthesis approach can generate control inputs such that the time-varying numbers of individuals in each category (e.g., infectious, immune) satisfy the MTL specifications.

Based on the simulation results, we observe that early control is essential in mitigating the spread of COVID-19, and more control effort is needed for more \textit{stringent} MTL specifications. For example, based on the model in Lombardy, Italy, achieving less than 100 deaths per day and 10000 total deaths within 100 days requires 441.7$\%$ more vaccination control effort than achieving less than 1000 deaths per day and 50000 total deaths within 100 days. As the control inputs are generated on a daily basis, the proposed approach can be used to assist and provide quantitative guidelines in public health control strategies to achieve specific specifications for mitigating the spread of COVID-19.

\section{Related Work}

\noindent\textbf{COVID-19 epidemic modeling and control strategies}: Ever since the outbreak of COVID-19, there has been numerous research focusing on the modeling of COVID-19 epidemic based on data collected from both the epicenters and other places \cite{Bertozzi_Franco2020,Chen2002,Carcione_model,Elie2020}. Among the various models, \textit{compartmental models} such as SEIR and SUQC models have been used frequently for the analysis of COVID-19. There has also been work in analyzing or predicting 
the spread of COVID-19 using artificial intelligence models \cite{Zheng_AI2020}, stochastic intensity models \cite{Chen2002}, etc. The models we use in this paper are based on the SEIR (both the standard and with shield immunity) and SUQC models, but we have replaced some essential parameters (e.g., the shield strength in shield immunity, the quarantine rate in quarantine control) with control inputs which can be synthesized to vary on a daily basis.

\noindent\textbf{Optimal control of epidemic models}: There exist work in designing vaccination control for the SEIR or SIR models of epidemics \cite{Alonso2012,Pinho2015}. However, such methods have not been applied in the setting of COVID-19. Besides, there has been no work in optimal control of epidemic models with formal specifications (e.g., expressed in temporal logic formulas).

\noindent\textbf{Control synthesis with temporal logic specifications}: There are three main categories of approaches to designing controllers that meet temporal logic specifications \cite{KHFP,RVFK2012,KKW2002,MH2019IFAC,STJC2010,UASX2011,liu2017distributed,Djeumou2020,liu2017communication,liu2020distributed,cubuktepe2020policy}. The first category of approaches abstract the system as a transition system and transform the control syntheses problem into a series of constrained reachability problems \cite{Topcu,Dimarogonas,Coogan}. The second category of approaches mainly focus on linear dynamical systems
and they convert the control synthesis problem into a mixed-integer linear programming (MILP) problem~\cite{BluSTL,Allerton2019,zheACC2019DF,sayan2016,zhe_advisory,zheACC2018wind} which can be solved efficiently by MILP solvers. The third category of approaches substitute the temporal logic constraint into the objective function of the optimization problem and apply a functional gradient descent algorithm on the resulting unconstrained problem \cite{Andygradient,Abbas2014,zhe_control,zheACCstorageControl}. The control synthesis approach in this paper is based on the second category of approaches, but we have extended the method to non-linear dynamical systems to fit the epidemic models for COVID-19.

% We first synthesize the vaccination control inputs according to the nominal system such that the nominal infection trajectory satisfies the MTL specification with pre-calculated robustness margins. Then we verify that all the possible infection trajectories within the uncertainty bounds satisfy the MTL specification. 

\section{Metric Temporal Logic (MTL)}        
\label{sec_MTL}   
In this section, we briefly review  metric temporal logic (MTL)~\cite{FainekosMTL} interpreted over discrete-time trajectories. The state $x$ (e.g., representing the susceptible, exposed, infectious, recovered population of a certain region) belongs to
the domain $\mathcal{X}\subset\mathbb{R}^n$. The time set is $\mathbb{T} = \mathbb{R}_{\ge0}$. The domain $\mathbb{B} = \{\textrm{True}, \textrm{False}\}$ is the Boolean domain, and the time index set is $\mathbb{I} = \{0,1,\dots\}$. We use $t[k]\in\mathbb{T}$ to denote the time instant at time index $k\in\mathbb{I}$ and $x[k]\triangleq x(t[k])$ to denote the value of $x$ at time $t[k]$. We use $\xi$ to denote a \textit{trajectory} as a function from $\mathbb{T}$ to $\mathcal{X}$. A set $AP$ is a set of atomic propositions, each mapping $\mathcal{X}$ to $\mathbb{B}$. The syntax of MTL is defined recursively as follows:
\[
\varphi:=\top\mid \pi\mid\lnot\varphi\mid\varphi_{1}\wedge\varphi_{2}\mid\varphi_{1}\vee
\varphi_{2}\mid\varphi_{1}\mathcal{U}_{\mathcal{I}}\varphi_{2},
\]
where $\top$ stands for the Boolean constant True, $\pi\in AP$ is an atomic
proposition, $\lnot$ (negation), $\wedge$ (conjunction), $\vee$ (disjunction)
are standard Boolean connectives, $\mathcal{U}$ is a temporal operator
representing \textquotedblleft until\textquotedblright, $\mathcal{I}$ is a time index interval of
the form $\mathcal{I}=[i_{1},i_{2}]$ ($i_1\le i_2$, $i_1, i_2\in\mathbb{I}$). We
can also derive two useful temporal operators from \textquotedblleft
until\textquotedblright~($\mathcal{U}$), which are \textquotedblleft
eventually\textquotedblright~$\Diamond_{\mathcal{I}}\varphi=\top\mathcal{U}_{\mathcal{I}}\varphi$ and
\textquotedblleft always\textquotedblright~$\Box_{\mathcal{I}}\varphi=\lnot\Diamond_{\mathcal{I}}\lnot\varphi$. For example, the MTL formula $\Box_{[0,100]} (DeathsPerDay\le 0.001)\wedge\Diamond_{[40, 60]}(Recovered\ge6)$ means ``the deaths from infection should never exceed 0.001 million (one thousand) per day within the next 100 days, and the immune population should eventually exceed 6 million after 40 to 60 days'' (we assume that the unit in $\pi$ is \textbf{million} and the unit in $\mathcal{I}$ is \textbf{day} in this paper, unless otherwise indicated).

% We define the set of states that satisfy the atomic proposition $\pi$ as $\mathcal{O}(\pi)\subset \mathcal{X}$. We denote the distance from $x$ to a set $D\subseteq\mathcal{X}$ as \textbf{dist}$_d(x,D)\triangleq$inf$\{d(x, x')\mid x'\in cl(D)\}$, where $d$ is a metric on $\mathcal{X}$ and $cl(D)$ denotes the closure of the set $D$. In this paper, we use the metric $d(x,x')=\norm{x-x'}$, where $\left\Vert\cdot\right\Vert $ denotes the 2-norm. We denote the depth of $x$ in $D$ as \textbf{depth}$_d(x,D)\triangleq$ \textbf{dist}$_d(x,\mathcal{X}\setminus D)$. We define the signed distance from $x$ to $D$ as $\textbf{Dist}_d(x,D)\triangleq-\textbf{dist}_d(x,D)$, if $x$ $\not\in D$; and $\textbf{Dist}_d(x,D)\triangleq\textbf{depth}_d(x,D)$, if $x$ $\in D$.
%\begin{equation}
%
%\begin{cases}
%-\textbf{dist}_d(x,D),& \mbox{if $x$ $\not\in D$},\\                      
%\textbf{depth}_d(x,D), & \mbox{if $x$ $\in D$}.
%\end{cases}                        
%\end{equation}

We define the set of states that satisfy the atomic proposition $\pi$ as $\mathcal{O}(\pi)\in \mathcal{X}$. We denote $\langle\langle\varphi\rangle\rangle(\xi,k)=\top$ if the trajectory $\xi$ satisfies the formula $\varphi$ at discrete-time
instants $t[k]$ ($k\in\mathbb{I}$). Then the Boolean semantics of MTL are defined recursively as follows~\cite{FAINEKOScontinous}:
	\[   
	\begin{split}
	\langle\langle\top\rangle\rangle(\xi,k) :=& \top,\\
	\langle\langle \pi\rangle\rangle(\xi,k)  :=& x[k]\in\mathcal{O}(\pi),\\
	\langle\langle \neg\varphi\rangle\rangle(\xi,k)  :=&\neg\langle\langle \varphi\rangle\rangle(\xi,k),\\
	\langle\langle\varphi_1\vee\varphi_2\rangle\rangle(\xi,k):=&\langle\langle\varphi_1\rangle\rangle(\xi,k)\vee\langle\langle\varphi_2\rangle\rangle(\xi,k),\\
	\langle\langle\varphi_1\mathcal{U}_{\mathcal{I}}\varphi_{2}\rangle\rangle(\xi,k)  :=&\bigvee_{k'\in (k+\mathcal{I})}\big(\langle\langle \varphi_2\rangle\rangle(\xi,k')\wedge\bigwedge_{k\le k''<k'}\langle\langle \varphi_1\rangle\rangle\\
	&(\xi,k'')\big), 
	\end{split}
	\]
	where $k+\mathcal{I}=\{k+\tilde{k}\vert \tilde{k}\in\mathcal{I}\}$.
	
% The Boolean semantics of MTL can be found in \cite{FAINEKOScontinous}, with the slight variation that we only evaluate the satisfaction of a trajectory with respect to an MTL formula at discrete-time instants $t[k]~(k\in\mathcal{I})$. The robustness degree of a trajectory $\xi$ with respect to an MTL formula $\varphi$ at time index $k$, denoted as $\left[\left[\varphi\right]\right](\xi, k)$, is defined recursively as follows:
% \[
% \begin{split}
% \left[\left[\top\right]\right](\xi, k) :=& +\infty,\\
% \left[\left[ \pi\right]\right](\xi, k)  :=&\textbf{Dist$_d(x[k],\mathcal{O}(\pi))$},\\
% \left[\left[ \neg\varphi\right]\right](\xi, k)  :=&-\left[\left[ \varphi\right]\right](\xi, k),\\
% \left[\left[\varphi_1\vee\varphi_2\right]\right](\xi, k)  :=&\max\big(\left[\left[ \varphi_1\right]\right](\xi, k),\left[\left[ \varphi_2\right]\right](\xi, k)\big),\\
% \left[\left[\varphi_1\mathcal{U}_{\mathcal{I}}\varphi_{2}\right]\right](\xi, k)  :=&\max_{k'\in (k+\mathcal{I})}\Big(\min\big(\left[\left[ \varphi_2\right]\right](\xi, k'),\\& \min_{k\leq k''<k'}\left[\left[\varphi_1\right]\right]
% (\xi,k'')\big)\Big).  
% \end{split} 
% \]
% As defined, $\left[\left[\varphi\right]\right](\xi, k)\ge0$ if $\xi$ satisfies $\varphi$ at time index $k$. 

\section{COVID-19 Models with Control Strategies}
\label{sec_models}
In this section, we study three models for COVID-19 epidemic  \cite{Carcione_model,Weitz_nature,Zhao2020} and introduce the corresponding models with vaccination control, shield immunity control and quarantine control.

\subsection{COVID-19 SEIR Model with Vaccination Control}
\begin{figure}
	\centering
	\includegraphics[scale=0.2]{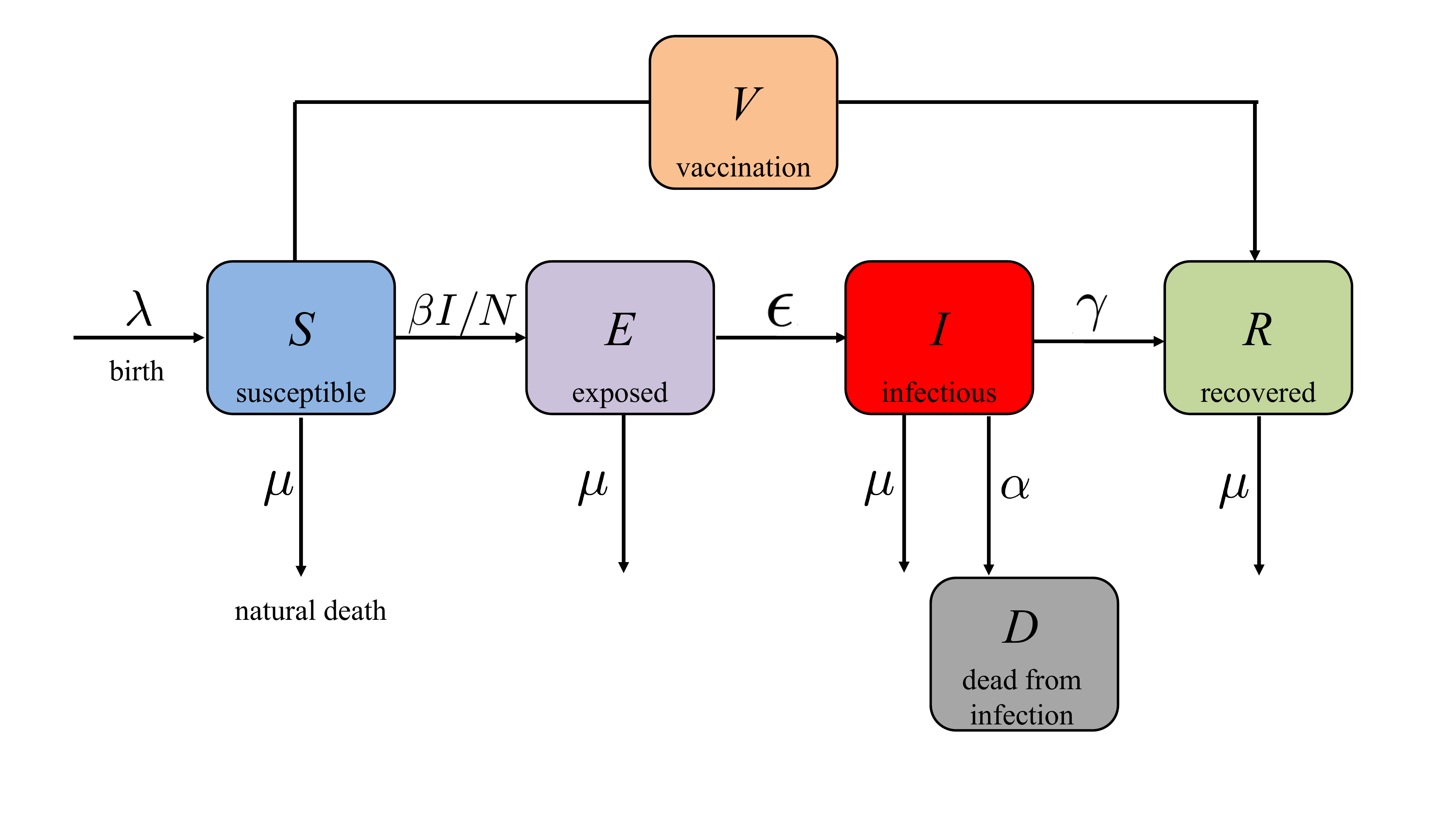}
	\caption{Block diagram of the COVID-19 SEIR model with vaccination control.}
	\label{diagram1}
\end{figure}

The susceptible, exposed, infectious, recovered (SEIR) model has been frequently used in epidemic analyses. As shown in Fig. \ref{diagram1}, the total population is divided into five subgroups: 
\begin{itemize}
    \item The susceptible population $S$: everyone is susceptible to the disease by birth since immunity is not hereditary; 
    \item The exposed population $E$: the individuals who have been exposed to the disease, but are still not infectious;
    \item The infectious population $I$: the individuals who are infectious;   
    \item The immune (recovered) population $R$: the individuals who are vaccinated or recovered from the disease, i.e., the population who are immune to the disease;
    \item The dead population $D$: the dead individuals from the disease.
\end{itemize}

We consider a COVID-19 SEIR model \cite{Carcione_model,Elie2020} with vaccination control \cite{Alonso2012} as follows. 
\begin{align}
\begin{split}
&\dot{I} = \epsilon E - (\gamma+\mu+\alpha)I;\\
&\dot{E} = \beta SI/N - (\mu+\epsilon)E;\\
&\dot{S} = \lambda N - \mu S - \beta SI/N - V;\\
&\dot{R} = \gamma I - \mu R + V;\\
&\dot{D} = - \dot{I}-\dot{E}-\dot{S}-\dot{R},
\end{split}            
\label{vaccination_model}
\end{align}
where the control input $V$ is the number of vaccinated individuals per day, $N = S + E + I + R\le N_0$ is the total population in the region ($N_0$ is the initial total population in the region), $S$, $E$, $I$, $R$ and $D$ are the number of susceptible, exposed, infectious and recovered population in the region, respectively, and $D$ is the number of deaths from SARS-CoV-2 in the region. For the parameters, $\lambda$ denotes the per-capita birth rate, 
$\mu$ is the per-capita natural death rate (death rate from causes unrelated to SARS-CoV-2), $\alpha$ is the SARS-CoV-2 virus-induced average fatality rate, $\beta$ is the probability of disease transmission per contact (dimensionless) times the number of contacts per unit
time, $\epsilon$ is the rate of progression from exposed to infectious (the
reciprocal is the incubation period), and $\gamma$ is the recovery rate of infectious individuals (the reciprocal is the
infectious period). Note that in (\ref{vaccination_model}), $D = N_0-I - E - S - R=N_0-N$ holds as we have assumed that the birth rate and the natural death rate are the same for the population we are investigating, i.e., $\lambda=\mu$.

\begin{remark}
Note that one difference between this model and the vaccination control model in \cite{Alonso2012} is that we control $V$ as the number of vaccinated individuals per day (constrained to be less than the susceptible population $S$), while in \cite{Alonso2012} the control input is the ratio of the vaccinated individuals per day to the average born population per day. We found it more convenient this way for computational convenience in the control synthesis in later sections.
\end{remark}

\subsection{COVID-19 SEIR Model with Shield Immunity Control}
\textit{Shield immunity} is a strategy recently proposed in \cite{Weitz_nature} to limit the transmission of SARS-CoV-2. The basic idea of this strategy is to
increase the proportion of interactions with recovered individuals as opposed to the other individuals in the population. The effectiveness of this strategy is based on the assumption that 
recovered individuals (virus-negative and antibody-positive) can
safely interact with both susceptible and infectious individuals without getting infected with the disease. 

\begin{figure}
	\centering
	\includegraphics[scale=0.2]{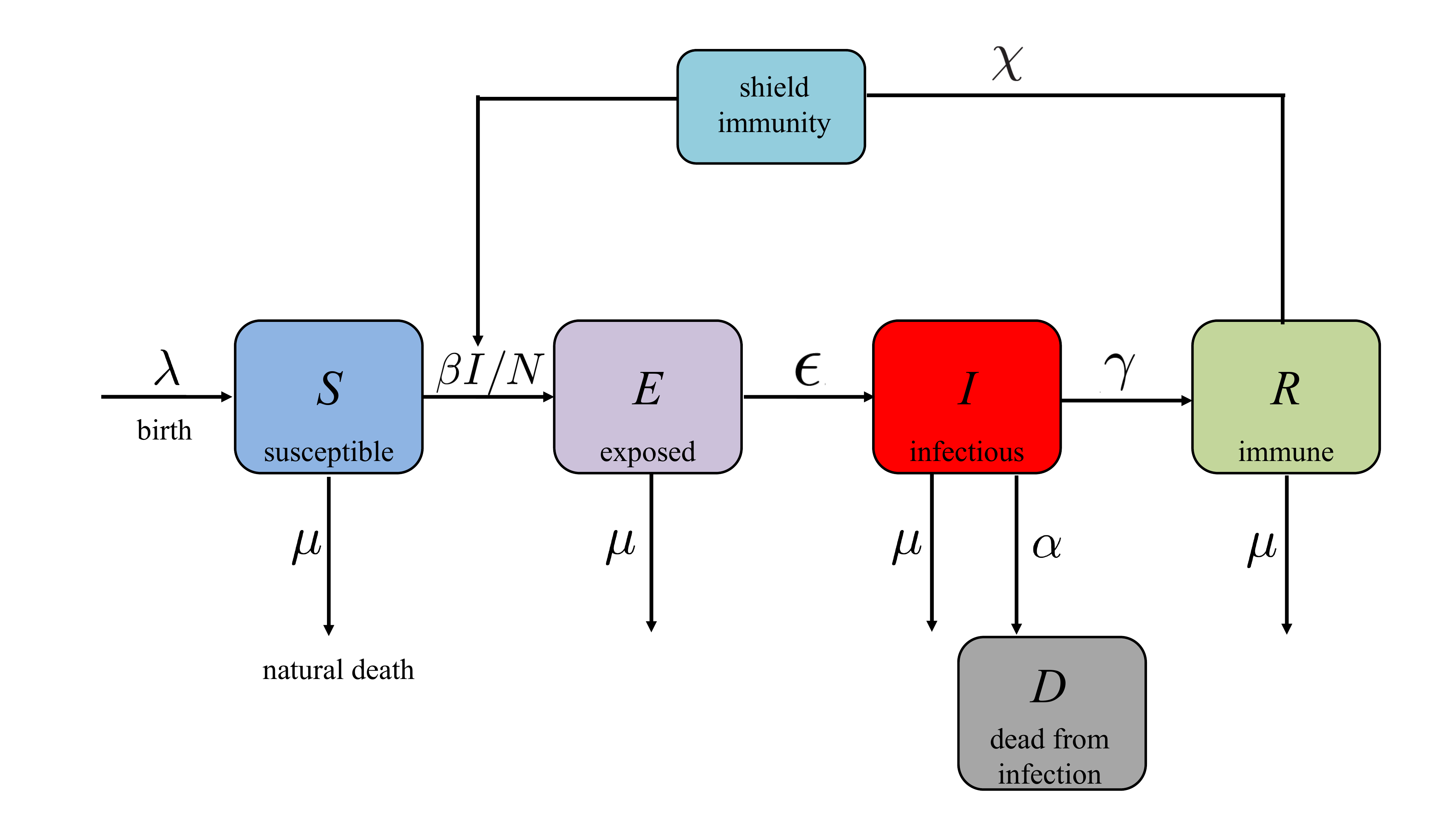}
	\caption{Block diagram of the COVID-19 SEIR model with shield immunity control.}
	\label{diagram2}
\end{figure}

As the model used in \cite{Weitz_nature} is modified from an SIR model, we consider a corresponding SEIR model with shield immunity control as follows (see Fig. \ref{diagram2} as an illustration).
\begin{align}
\begin{split}
& \dot{I} = \epsilon E - (\gamma+\mu+\alpha)I ;\\
& \dot{E} = \beta SI/(N+\chi R) - (\mu+\epsilon)E;\\
& \dot{S} = \lambda N - \mu S - \beta SI/(N+\chi R);\\
& \dot{R} = \gamma I - \mu R;\\
& \dot{D} = - \dot{I}-\dot{E}-\dot{S}-\dot{R},
\end{split}      
\label{shield_model}
\end{align}
where the states and parameters are the same as in (\ref{vaccination_model}), while $\chi(\cdot)$ is the \textit{shield strength} \cite{Weitz_nature} as control input to be synthesized for the recovered population to substitute the contact for the susceptible population.

\subsection{COVID-19 SUQC Model with Quarantine Control}
\begin{figure}
	\centering
	\includegraphics[scale=0.2]{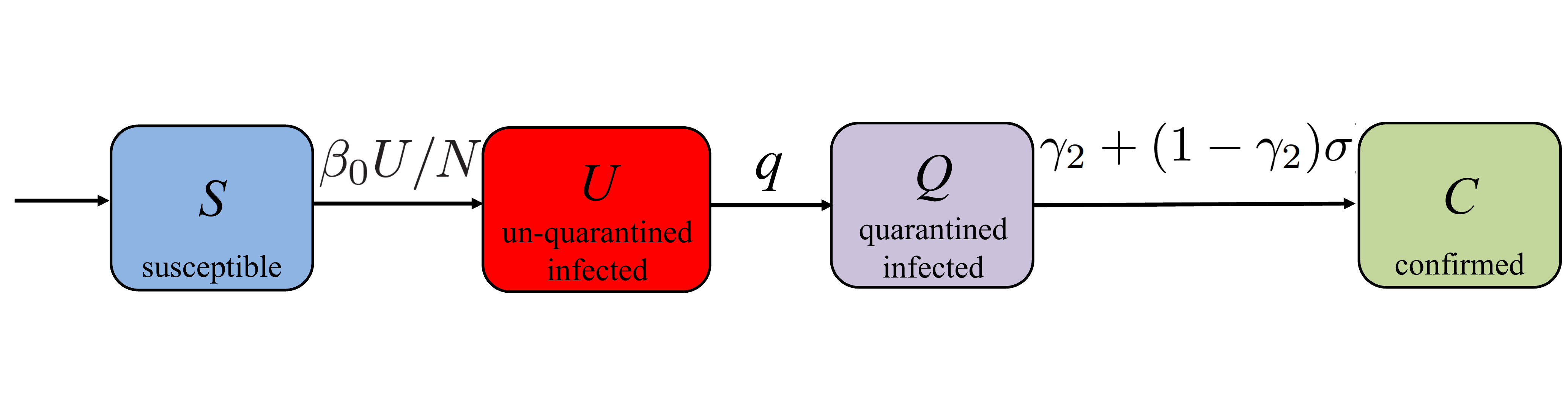}
	\caption{Block diagram of the COVID-19 SUQC model with quarantine control.}
	\label{diagram3}
\end{figure}

The susceptible, un-quarantined infected, quarantined infected, confirmed infected (SUQC) model was recently proposed in \cite{Zhao2020} based on the COVID-19 data in Wuhan, China. As shown in Fig. \ref{diagram3}, we consider four subgroups in the population:
\begin{itemize}
    \item The susceptible population $S$: everyone is susceptible to the disease by birth since immunity is not hereditary; 
    \item The un-quarantined infected population $U$: the individuals who are infected and un-quarantined, and they can be either asymptomatic or symptomatic;
    \item The quarantined infected population $Q$: the individuals who are infectious and quarantined (the un-quarantined infected become quarantined infected by isolation or hospitalization, and the quarantined infected lose the ability of infecting the susceptible individuals);   
    \item The confirmed infected population $C$: the individuals who are confirmed to be infected with the disease (i.e., the positive cases).
    \end{itemize}

We consider the SUQC model with quarantine control as follows.
\begin{align}
\begin{split}
&\dot{S} = - \beta_0 US/N;\\
&\dot{U} = \beta_0 US/N - q U;\\
&\dot{Q} = q U - (\gamma_2 + (1-\gamma_2)\sigma)Q;\\
&\dot{C} = (\gamma_2 + (1-\gamma_2)\sigma)Q,
\end{split}         
\label{quarantine_model}
\end{align}
where $q$ is the quarantine rate (for an un-quarantined infected to be quarantined) as control input to be synthesized, $S$, $U$, $Q$ and $C$ are the number of susceptible, un-quarantined infected, quarantined infected and confirmed infected population in the region, respectively, $\beta_0$ is the infection rate (i.e., the mean number of new infected caused by an un-quarantined infected per day), $\gamma_2$ is the confirmation rate of $Q$ (i.e., the probability that the
quarantined infected are identified to be confirmed
cases through conventional methods such as laboratory
diagnosis), $\sigma$ is the subsequent confirmation rate of those infected that are not confirmed by the conventional methods, but
confirmed with additional tests.

% $\beta_0 U/N$

\section{Control Synthesis of COVID-19 Epidemic with Metric Temporal Logic Specifications}
\label{sec_control}
In this section, we present the control synthesis methods for the three COVID-19 epidemic models in Section \ref{sec_models} with vaccination control, shield immunity control and quarantine control, respectively.\\

\noindent\textbf{Vaccination control}: For the COVID-19 SEIR model with vaccination control, we discretize the model in (\ref{vaccination_model}) as follows. 
\begin{align}\nonumber
\begin{split}
& I[k+1] = I[k] + T_{\textrm{s}}\epsilon E[k] - T_{\textrm{s}}(\gamma+\mu+\alpha)I[k];\\
& E[k+1] = E[k] + T_{\textrm{s}}\beta S[k]I[k]/N[k] - T_{\textrm{s}}(\mu+\epsilon)E[k];
\end{split} 
\end{align}
\begin{align}
\begin{split}
& S[k+1] = S[k] + T_{\textrm{s}}\lambda N[k] - T_{\textrm{s}}\mu S[k] - T_{\textrm{s}}\beta S[k]I[k]/N[k] \\& ~~~~~~~~~~~~~ - T_{\textrm{s}}V[k];\\
& R[k+1] = R[k] + T_{\textrm{s}}\gamma I[k] - T_{\textrm{s}}\mu R[k] + T_{\textrm{s}}V[k];\\
& D[k] = N_0 - I[k] - E[k] - S[k] -R[k],
\end{split}    
\label{discrete_vaccination_model}
\end{align}
where $T_{\textrm{s}}$ is the sampling period. We also use $\Delta D[k]=D[k]-D[k-1]$ to denote the number of deaths from the infection at day $k$. 

Following the notations in Section \ref{sec_MTL}, we use $x_{\textrm{V}}=[I, E, S, R, D]$ to denote the state of (\ref{discrete_vaccination_model}) and $\xi^{\textrm{V}}_{\bm\cdot;x^{init}_{\textrm{V}},V}$ to denote the trajectory of (\ref{discrete_vaccination_model}) starting from $x^{init}_{\textrm{V}}=[I[0], E[0], S[0], R[0], D[0]]$ and vaccination control input $V[\bm\cdot]$. 

% We formulate the problem of vaccination control as follows.
\begin{problem} [Vaccination control]
	Given the SEIR model with vaccination control in (\ref{discrete_vaccination_model}) and an MTL specification $\varphi_{\textrm{V}}$, compute the control input $V[\bm\cdot]$ that minimizes the vaccination control efforts $\norm{V[\bm\cdot]}$ while satisfying $\langle\langle\varphi_{\textrm{V}}\rangle\rangle(\xi^{\textrm{V}}_{\bm\cdot;x^{init}_{\textrm{V}},V}, 0)=\top$, i.e., the trajectory $\xi^{\textrm{V}}_{\bm\cdot;x^{init}_{\textrm{V}},V}$ satisfies the MTL specification $\varphi_{\textrm{V}}$. 
	\label{problem1}                               
\end{problem}     

The vaccination control synthesis problem can be formulated as a constrained optimization problem as follows.
\begin{align}
\begin{split}
\underset{V[\bm\cdot]}{\min} ~ & \norm{V[\bm\cdot]}   \\
\text{s.t.} ~ 
& I[k+1] = I[k] + T_{\textrm{s}}\epsilon E[k] - T_{\textrm{s}}(\gamma+\mu+\alpha)I[k], \\&~~~~~~~~~~~~~\forall k=0,\dots,T-1, \\
& E[k+1] = E[k] + T_{\textrm{s}}\beta S[k]I[k]/N[k] - T_{\textrm{s}}(\mu+\epsilon)E[k], \\&~~~~~~~~~~~~~\forall k=0,\dots,T-1,  \\
& S[k+1] = S[k] + T_{\textrm{s}}\lambda N[k]- T_{\textrm{s}}\mu S[k] - T_{\textrm{s}}\beta S[k]I[k]/N[k]\\&~~~~~~~~~~~~~  - T_{\textrm{s}}V[k], \forall k=0,\dots,T-1, \\
& R[k+1] = R[k] + T_{\textrm{s}}\gamma I[k] - T_{\textrm{s}}\mu R[k] + T_{\textrm{s}}V[k], \\&~~~~~~~~~~~~~ \forall k=0,\dots,T-1, \\
& D[k] = N_0 - I[k] - E[k] - S[k] -R[k], \forall k=0,\dots,T,  \\
& 0\le V[k]\le S[k], \forall k=0,\dots,T, \nonumber \\ 
& \langle\langle\varphi_{\textrm{V}}\rangle\rangle(\xi^{\textrm{V}}_{\bm\cdot;x^{init}_{\textrm{V}},V}, 0)=\top, 
\end{split}
\end{align} 
where $T\in\mathbb{I}$ is the maximal time index we consider.

The above optimization problem is generally a mixed-integer non-linear programming problem. We refer the readers to \cite{sayan2016} for a detailed description of how the constraint $\langle\langle\varphi_{\textrm{V}}\rangle\rangle(\xi^{\textrm{V}}_{\bm\cdot;x^{init}_{\textrm{V}},V}, 0)=\top$ is encoded to satisfy an MTL specification $\varphi_{\textrm{V}}$. The integer variables are introduced when a \textit{big-M formulation} \cite{Schrijver86} is needed to satisfy MTL specifications such as $\Diamond_{[0,10]}\varphi$ ($\varphi$ should hold true for at least one day during the first 10 days) or $\varphi_1\vee\varphi_2$ (at least one of the MTL formulas $\varphi_1$, $\varphi_2$ should hold true). As the change of total population is relatively small compared to the multiplication of the susceptible population and the infectious population, we approximate the term $T_{\textrm{s}}\beta S[k]I[k]/N[k]$ with $T_{\textrm{s}}\beta S[k]I[k]/N_0$. With such an approximation, the optimization problem becomes a mixed-integer bi-linear programming problem, which can be more efficiently solved using techniques such as McCormick's relaxation \cite{McCormick1976,Gupte2013SolvingMI}. Furthermore, if the MTL specification $\varphi$ consists of only conjunctions ($\wedge$) and the always operator ($\Box$), the integers in the optimization problem can be eliminated \cite{sayan2016} and the problem becomes a bi-linear programming problem.\\

\noindent\textbf{Shield immunity control}: For the COVID-19 SEIR model with shield immunity control, we discretize the model in (\ref{shield_model}) as follows. 
\begin{align}
\begin{split}
& I[k+1] = I[k] + T_{\textrm{s}}\epsilon E[k] - T_{\textrm{s}}(\gamma+\mu+\alpha)I[k];\\
& E[k+1] = E[k] + T_{\textrm{s}}\beta S[k]I[k]/(N[k]+\chi[k]R[k]) \\& ~~~~~~~~~~~~~ - T_{\textrm{s}}(\mu+\epsilon)E[k];\\
& S[k+1] = S[k] + T_{\textrm{s}}\lambda N[k] - T_{\textrm{s}}\mu S[k] - T_{\textrm{s}}\beta S[k]I[k]/(N[k]\\&~~~~~~~~~~~~~ +\chi[k]R[k]);\\
& R[k+1] = R[k] + T_{\textrm{s}}\gamma I[k] - T_{\textrm{s}}\mu R[k];\\
& D[k] = N_0 - I[k] - E[k] - S[k] -R[k],
\end{split}    
\label{discrete_shield_model}
\end{align}
where $T_{\textrm{s}}$ is the sampling period.\\

Following the notations in Section \ref{sec_MTL}, we use $x_{\textrm{S}}=[I, E, S, R, D]$ to denote the state of (\ref{discrete_shield_model}) and $\xi^{\textrm{S}}_{\bm\cdot;x^{init}_{\textrm{S}},\chi}$ to denote the trajectory of (\ref{discrete_shield_model}) starting from $x^{init}_{\textrm{S}}=[I[0], E[0], S[0], R[0], D[0]]$ and shield immunity control input $\chi[\bm\cdot]$. 

% We formulate the problem of shield immunity control as follows.
\begin{problem} [Shield immunity control]
	Given the SEIR model with shield immunity control in (\ref{discrete_shield_model}) and an MTL specification $\varphi_{\textrm{S}}$, compute the control input $\chi[\bm\cdot]$ that minimizes the shield immunity control efforts $\norm{\chi[\bm\cdot]}$ while satisfying $\langle\langle\varphi_{\textrm{S}}\rangle\rangle(\xi^{\textrm{S}}_{\bm\cdot;x^{init}_{\textrm{S}},\chi}, 0)=\top$, i.e., the trajectory $\xi^{\textrm{S}}_{\bm\cdot;x^{init}_{\textrm{S}},\chi}$ satisfies the MTL specification $\varphi_{\textrm{S}}$. 
	\label{problem1}                               
\end{problem}  

The shield immunity control synthesis problem can be formulated as a constrained optimization problem as follows.
\begin{align}
\begin{split}
\underset{\chi[\bm\cdot]}{\min} ~ & \norm{\chi[\bm\cdot]}   \\
\text{s.t.} ~ 
& I[k+1] = I[k] + T_{\textrm{s}}\epsilon E[k] - T_{\textrm{s}}(\gamma+\mu+\alpha)I[k], \\&~~~~~~~~~~~~~ \forall k=0,\dots,T-1, \\
& E[k+1] = E[k] + T_{\textrm{s}}\beta S[k]I[k]/(N[k]+\chi[k]R[k]) \\&~~~~~~~~~~~~~ - T_{\textrm{s}}(\mu+\epsilon)E[k], \forall k=0,\dots,T-1,  \\
& S[k+1] = S[k] + T_{\textrm{s}}\lambda N[k]- T_{\textrm{s}}\mu S[k]- T_{\textrm{s}}\beta S[k]\\&~~~~~~~~~~~~~   \times I[k]/(N[k]+\chi[k]R[k]), \forall k=0,\dots,T-1, \\
& R[k+1] = R[k] + T_{\textrm{s}}\gamma I[k] - T_{\textrm{s}}\mu R[k], \forall k=0,\dots,T-1, \\
& D[k] = N_0 - I[k] - E[k] - S[k] -R[k], \forall k=0,\dots,T,  \\
& 0\le \chi[k]\le \chi_{\textrm{max}}, \forall k=0,\dots,T, \nonumber \\ 
& \langle\langle\varphi_{\textrm{S}}\rangle\rangle(\xi^{\textrm{S}}_{\bm\cdot;x^{init}_{\textrm{S}},\chi}, 0)=\top, 
\end{split}
\end{align} 
where $T\in\mathbb{I}$ is the maximal time index we consider, and $\chi_{\textrm{max}}$ is the maximal shield strength.

The above optimization problem is generally a mixed-integer fractional constrained programming problem. If the MTL specification $\varphi$ consists of only conjunctions ($\wedge$) and the always operator ($\Box$), the integers in the optimization problem can be eliminated \cite{sayan2016} and the problem becomes a fractional constrained programming problem.\\

\noindent\textbf{Quarantine control}: For the COVID-19 SUQC model with quarantine control, we discretize the model in (\ref{quarantine_model}) as follows. 
\begin{align}
\begin{split}
& S[k+1] = S[k] - T_{\textrm{s}}\beta_0 U[k]S[k]/N[k];\\
& U[k+1] = U[k] + T_{\textrm{s}}\beta_0 U[k]S[k]/N[k] - q[k]U[k];\\
& Q[k+1] = Q[k] + T_{\textrm{s}}q[k]U[k] - T_{\textrm{s}}(\gamma_2 + (1-\gamma_2)\sigma)Q[k];\\
& C[k+1] = C[k] + T_{\textrm{s}}(\gamma_2 + (1-\gamma_2)\sigma)Q[k],\\
\end{split}     
\label{discrete_quarantine_model}
\end{align}
where $T_{\textrm{s}}$ is the sampling period. We also use $\Delta C[k]=C[k]-C[k-1]$ to denote the number of confirmed infected individuals at day $k$.

Following the notations in Section \ref{sec_MTL}, we use $x_{\textrm{Q}}=[S, U, Q, C]$ to denote the state of (\ref{discrete_quarantine_model}) and $\xi^{\textrm{Q}}_{\bm\cdot;x^{init}_{\textrm{Q}},q}$ to denote the trajectory of (\ref{discrete_quarantine_model}) starting from $x^{init}_{\textrm{Q}}=[S[0], U[0], Q[0], C[0]]$ and quarantine control input $q[\bm\cdot]$. 

% We formulate the problem of quarantine control as follows.
\begin{problem} [Quarantine control]
	Given the SUQC model with quarantine control in (\ref{discrete_quarantine_model}) and an MTL specification $\varphi_{\textrm{Q}}$, compute the control input $q[\bm\cdot]$ that minimizes the quarantine control efforts $\norm{q[\bm\cdot]}$ while satisfying $\langle\langle\varphi_{\textrm{Q}}\rangle\rangle(\xi^{\textrm{Q}}_{\bm\cdot;x^{init}_{\textrm{Q}},q}, 0)=\top$, i.e., the trajectory $\xi^{\textrm{Q}}_{\bm\cdot;x^{init}_{\textrm{Q}},q}$ satisfies the MTL specification $\varphi_{\textrm{Q}}$. 
	\label{problem1}                                          
\end{problem}  

The quarantine control synthesis problem can be formulated as a constrained optimization problem as follows.
\begin{align}
\begin{split}
\underset{q[\bm\cdot]}{\min} ~ & \norm{q[\bm\cdot]}  \\
\text{s.t.} ~ 
& S[k+1] = S[k] - T_{\textrm{s}}\beta_0 U[k]S[k]/N[k], \forall k=0,\dots,T-1, \\
& U[k+1] = U[k] + T_{\textrm{s}}\beta_0 U[k]S[k]/N[k] - q[k]U[k], \\&~~~~~~~~~~~~~\forall k=0,\dots,T-1,  \\
& Q[k+1] = Q[k] + T_{\textrm{s}}q[k]U[k] - T_{\textrm{s}}(\gamma_2 + (1-\gamma_2)\sigma)Q[k], \\&~~~~~~~~~~~~~\forall k=0,\dots,T-1, \\
& C[k+1] = C[k] + T_{\textrm{s}}(\gamma_2 + (1-\gamma_2)\sigma)Q[k], \\&~~~~~~~~~~~~~\forall k=0,\dots,T-1,   \\
& 0\le q[k]\le q_{\textrm{max}}, \forall k=0,\dots,T, \nonumber \\ 
& \langle\langle\varphi_{\textrm{Q}}\rangle\rangle(\xi^{\textrm{Q}}_{\bm\cdot;x^{init}_{\textrm{Q}},q}, 0)=\top, 
\end{split}
\end{align} 
where $T\in\mathbb{I}$ is the maximal time index we consider, and $q_{\textrm{max}}$ is the maximal quarantine rate.

The above optimization problem is generally a mixed-integer non-linear programming problem. As the change of total population is relatively small compared to the multiplication of the susceptible population and the un-quarantined infectious population, we approximate the term $T_{\textrm{s}}\beta_0 U[k]S[k]/N[k]$ with $T_{\textrm{s}}\beta_0 U[k]S[k]/\hat{N}_0$ (we use $\hat{N}_0$ to denote the initial population in the region in the scenario with quarantine control). With such an approximation, the optimization problem becomes a mixed-integer bi-linear programming problem, which can be more efficiently solved using techniques such as McCormick's relaxation \cite{McCormick1976,Gupte2013SolvingMI}. Furthermore, if the MTL specification $\varphi$ consists of only conjunctions ($\wedge$) and the always operator ($\Box$), the integers in the optimization problem can be eliminated \cite{sayan2016} and the problem becomes a bi-linear programming problem.\\

% \section{Verification of Vaccination Control Against Variations}

% \section{Optimal Vaccination Control}
% We consider the following Lagrangian:
% \begin{align}
% \begin{split}
% L(x,u,q,p_1,p_2,r)=&1/2\sum_{t=1}^{H}V(t)^2 + \sum_{t=1}^{H-1}q(t+1)^T\Big(f(x(t))x(t)+BV(t)-x(t+1)\Big) \\&+ p_1(t)(-V(t)) + p_2(t)(V(t)-V0) + r(t)\sum_{t=k}^{H-1}(Ax(t)-C),
% \end{split}
% \end{align}

% According to the KKT conditions, we have
% \begin{align}
% \begin{split}
% \frac{\partial L}{\partial x(t)}=&q^{\ast}(t+1)^{T}\Big(\frac{\partial f(x(t))}{\partial x(t)}x(t)+f(x(t))\Big)-q^{\ast}(t)^{T}+ r(t)A=0, t=1,2,\dots, H-1,
% \\ \frac{\partial L}{\partial x(H)}=&-q^{\ast}(H)^{T}=0,
% \\ \frac{\partial L}{\partial V(t)}=& V(t)+q^{\ast}(t+1)^{T}B-p_1(t)+p_2(t)=0, t=1,2,\dots, H-1,
% \end{split}
% \end{align}

\section{Simulation Results} 
In this section, we implement the proposed control synthesis approach in the three different control models as introduced in Section \ref{sec_models}.

\subsection{COVID-19 SEIR Model with Vaccination Control}
\label{results_vaccination}
\begin{figure}
	\centering
	\begin{subfigure}[b]{0.24\textwidth}
		\centering
		\includegraphics[width=\textwidth]{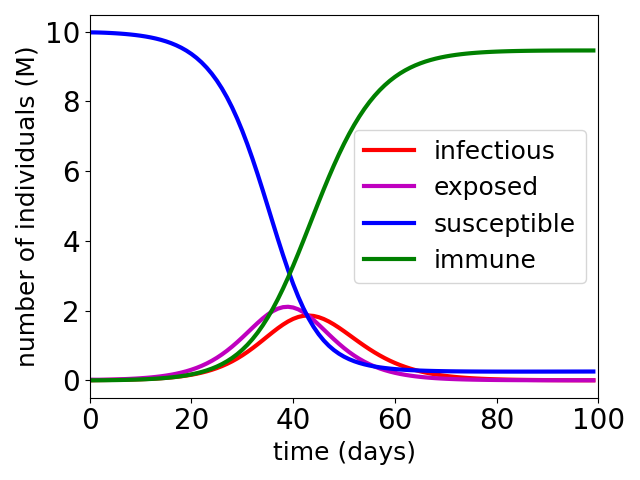}
		\caption{Number of individuals}
	\end{subfigure} 
		\hfill
	\begin{subfigure}[b]{0.24\textwidth}
		\centering
		\includegraphics[width=\textwidth]{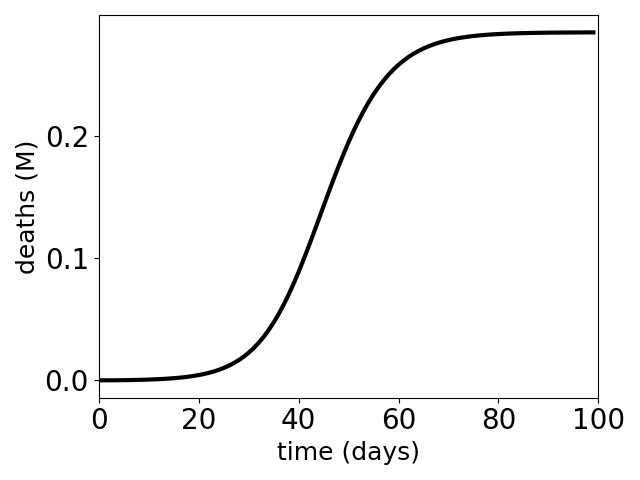}
		\caption{Number of deaths}
	\end{subfigure}
			\hfill
	\begin{subfigure}[b]{0.24\textwidth}
		\centering
		\includegraphics[width=\textwidth]{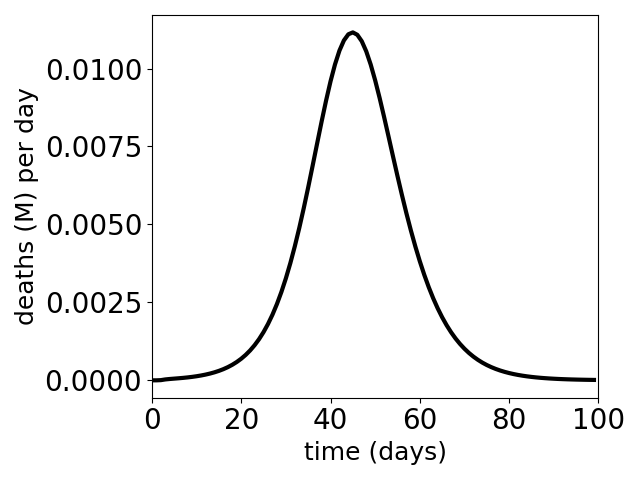}
		\caption{Number of deaths per day}
	\end{subfigure}
	\caption{Simulation results for COVID-19 SEIR model estimated from data from Lombardy, Italy with no isolation measures.}  
	\label{fig_results0}
\end{figure}

The parameters of the COVID-19 SEIR model are shown in Table \ref{parameter_SEIR}. They were estimated in \cite{Carcione_model} from the data in the early days (from February 23 to March 16, 2020) in Lombardy, Italy with no isolation measures. The start time for the simulations in this subsection are February 23, 2020. We consider three MTL specifications as shown in Table \ref{result_vaccination}. For example, $\varphi_{\textrm{V}}^1=\Box_{[0,100]} (\Delta D\le 0.001)\wedge\Box_{[0,100]} (D\le0.05)\wedge\Diamond_{[40, 60]}(R\ge6)$, which means ``the deaths from infection should never exceed 0.001 million (i.e., 
one thousand) per day and 0.05 million (i.e., 50 thousand) in total within the next 100 days, and the immune population should eventually exceed 6 million after 40 to 60 days''. We choose the initial values of the states as $I[0]=1000$ (people), $E[0]=0.02$ million, $S[0]=9.979$ million, $R[0]=0$ and $D[0]=0$, with $S[0] + E[0] + I[0] + R[0]+D[0]=N_0=10$ million. Fig. \ref{fig_results0} shows the simulation results without any vaccination. It can be seen that the three MTL specifications $\varphi_{\textrm{V}}^1$, $\varphi_{\textrm{V}}^2$ and $\varphi_{\textrm{V}}^3$ are all violated in such a situation. Note that as isolation measures (i.e., home isolation, social distancing and partial national lockdown) were taken since March 16 in Lombardy, Italy, the real situation was better than those shown in Fig. \ref{fig_results0}. Now we investigate the hypothetical scenario where the isolation measures are replaced by vaccination.

	\begin{table}[]
		\centering
		\caption{Parameters of COVID-19 SEIR model estimated from data from Lombardy, Italy from February 23 to March 16 (2020) with no isolation measures \cite{Carcione_model}.}
		\label{parameter_SEIR}  
		\begin{tabular}{llll}
			\toprule[2pt]    
			parameter    & value & parameter    & value\\ \hline
		    ~~~~$\lambda$        & 1/30295 & ~~~~$\epsilon$        & 0.2/day \\ 
		    ~~~~$\mu$        & 1/30295  & ~~~~$\gamma$        & 0.2/day\\ 
		    ~~~~$\alpha$        & 0.006/day &  ~~~~$N_0$         & 10 million \\ 
		    ~~~~$\beta$        & 0.75/day & ~~~~$T_{\textrm{s}}$  & 1 day
            \\ \bottomrule[2pt]       
		\end{tabular}
	\end{table}    
	
		\begin{table}[]
		\centering
		\caption{MTL specifications and simulation results for vaccination control (Section \ref{results_vaccination}).}
		\label{result_vaccination}  
		\begin{tabular}{ll>{\raggedright\arraybackslash}p{20mm}}
		\toprule[2pt]    
			 ~~~~~MTL specification & \tabincell{c}{control\\ effort} & \tabincell{c}{computation\\ time}\\ \hline
		    \tabincell{c}{$\varphi_{\textrm{V}}^1=\Box_{[0,100]} (\Delta D\le0.001)$\\ $~~~~~~\wedge\Box_{[0,100]} (D\le0.05)$\\ $~~~\wedge\Diamond_{[40, 60]}(R\ge6)$}   & ~1.28 & ~~~1.365 s  \\ 
		     \tabincell{c}{$\varphi_{\textrm{V}}^2=\Box_{[0,100]} (\Delta D\le0.0005)$\\ $~~~~~\wedge\Box_{[0,100]} (D\le0.02)$\\
		     $~~\wedge\Diamond_{[40, 60]}(R\ge6)$}  & ~1.927 & ~~~2.276 s \\ 
		     \tabincell{c}{$\varphi_{\textrm{V}}^3=\Box_{[0,100]} (\Delta D\le0.0001)$\\ $~~~~~\wedge\Box_{[0,100]} (D\le0.01)$\\$\wedge\Diamond_{[40, 60]}(R\ge6)$} & ~6.934 & ~~~3.289 s \\ \bottomrule[2pt]       
		\end{tabular}
	\end{table}  

\begin{figure*}
	\centering
	\begin{subfigure}[b]{0.24\textwidth}
		\centering
		\includegraphics[width=\textwidth]{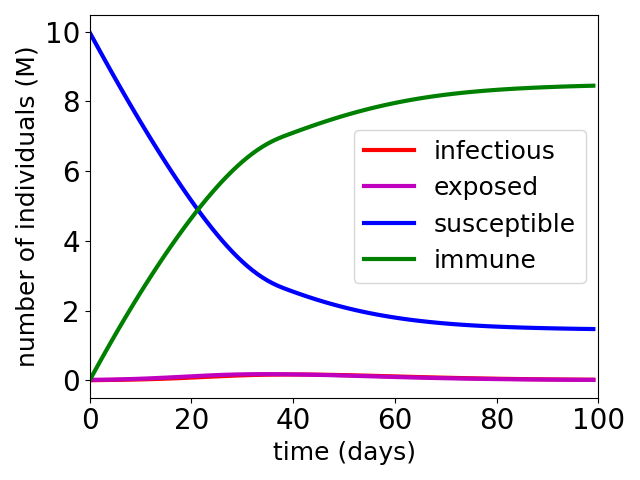}
%		\caption{Number of individuals}
	\end{subfigure} 
	\hfill
	\begin{subfigure}[b]{0.24\textwidth}
		\centering
		\includegraphics[width=\textwidth]{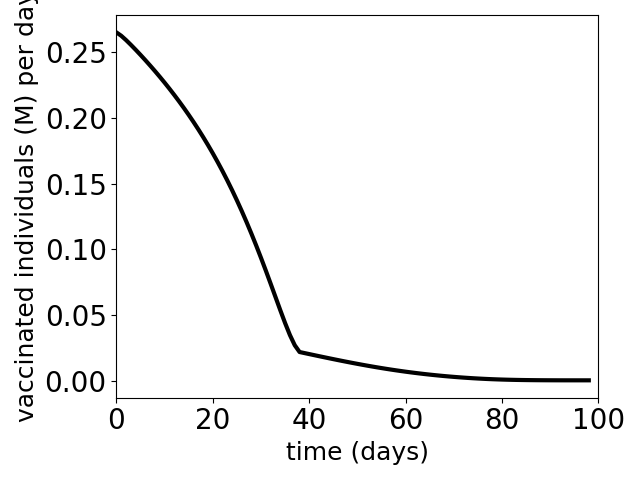}
%		\caption{Vaccinated individuals per day}
	\end{subfigure}
		\hfill
	\begin{subfigure}[b]{0.24\textwidth}
		\centering
		\includegraphics[width=\textwidth]{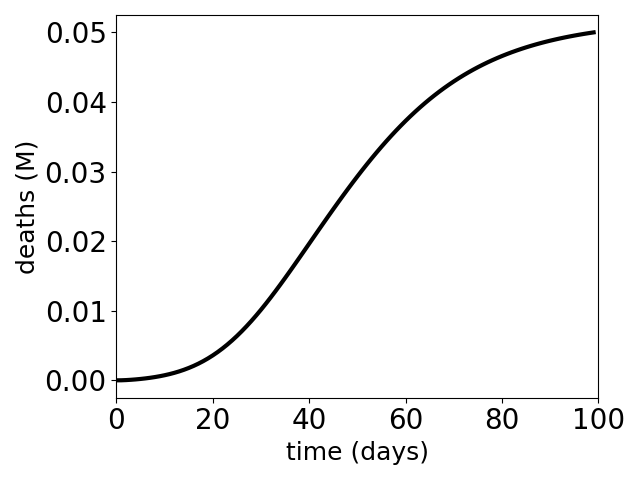}
%		\caption{Number of deaths}
	\end{subfigure}
			\hfill
	\begin{subfigure}[b]{0.24\textwidth}
		\centering
		\includegraphics[width=\textwidth]{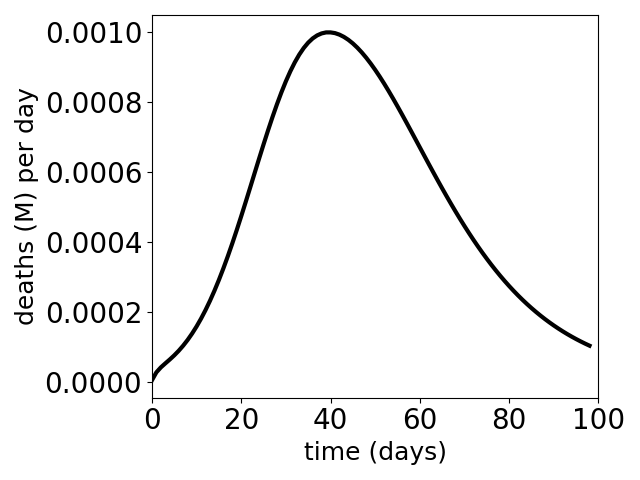}
%		\caption{Number of deaths per day}
	\end{subfigure}
\\
	\begin{subfigure}[b]{0.24\textwidth}
		\centering
		\includegraphics[width=\textwidth]{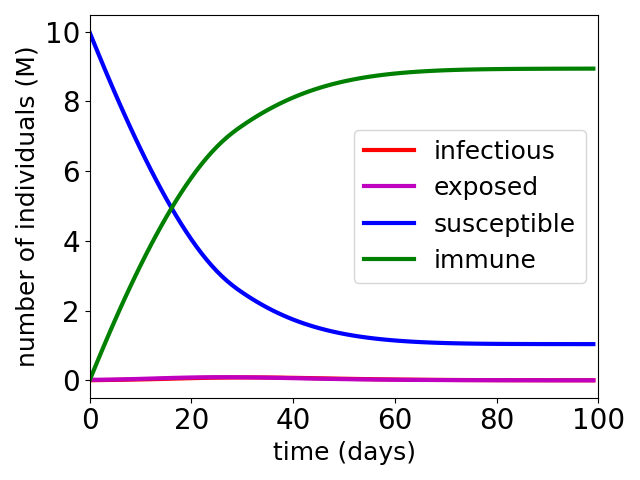}
%		\caption{Number of individuals}
	\end{subfigure}
	\hfill
	\begin{subfigure}[b]{0.24\textwidth}
		\centering
		\includegraphics[width=\textwidth]{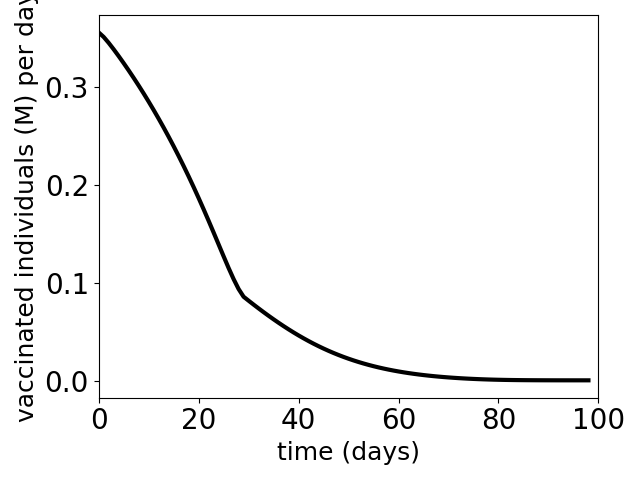}
%		\caption{Vaccinated individuals per day}
	\end{subfigure}
		\hfill
	\begin{subfigure}[b]{0.24\textwidth}
		\centering
		\includegraphics[width=\textwidth]{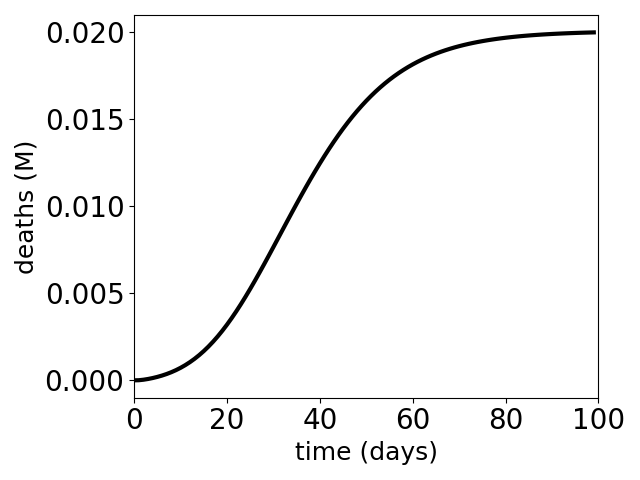}
%		\caption{Number of deaths}
	\end{subfigure}
			\hfill
	\begin{subfigure}[b]{0.24\textwidth}
		\centering
		\includegraphics[width=\textwidth]{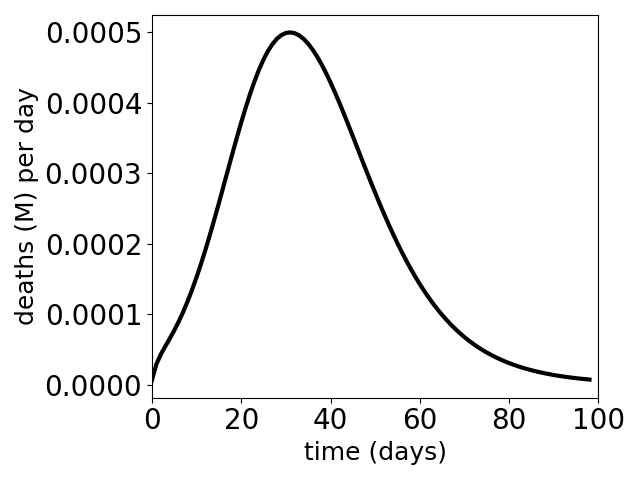}
%		\caption{Number of deaths per day}
	\end{subfigure}
	\\
	\begin{subfigure}[b]{0.24\textwidth}
		\centering
		\includegraphics[width=\textwidth]{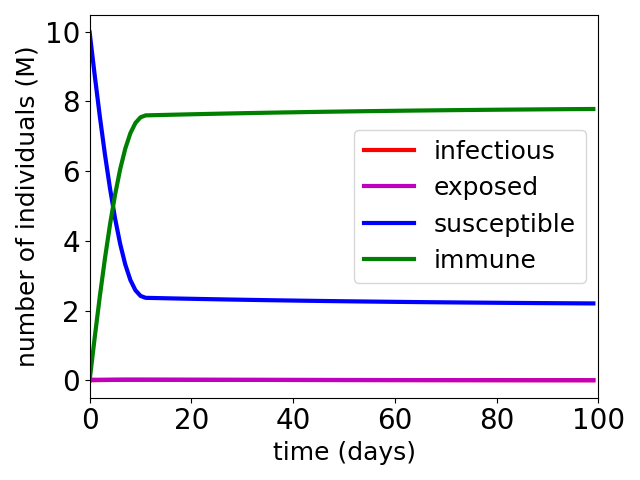}
		\caption{Number of individuals}
	\end{subfigure} 
	\hfill
	\begin{subfigure}[b]{0.24\textwidth}
		\centering
		\includegraphics[width=\textwidth]{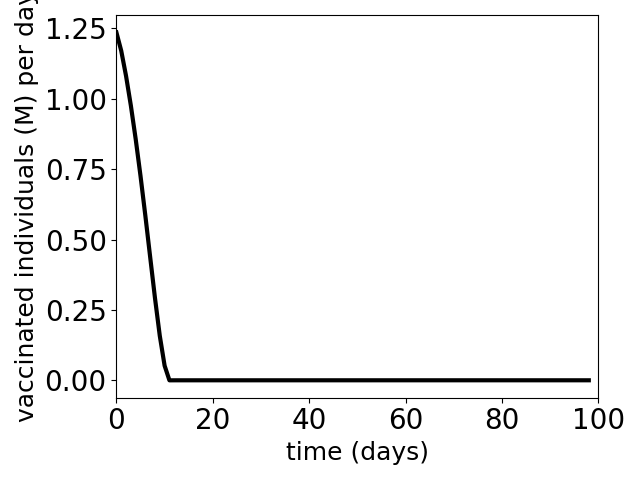}
		\caption{Vaccinated individuals per day}
	\end{subfigure}
		\hfill
	\begin{subfigure}[b]{0.24\textwidth}
		\centering
		\includegraphics[width=\textwidth]{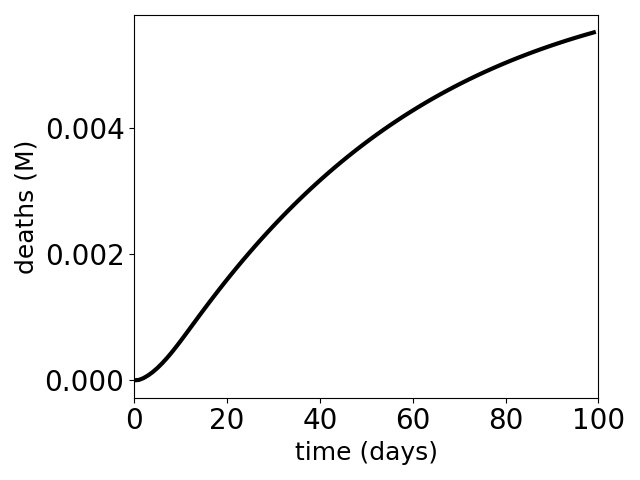}
		\caption{Number of deaths}
	\end{subfigure}
			\hfill
	\begin{subfigure}[b]{0.24\textwidth}
		\centering
		\includegraphics[width=\textwidth]{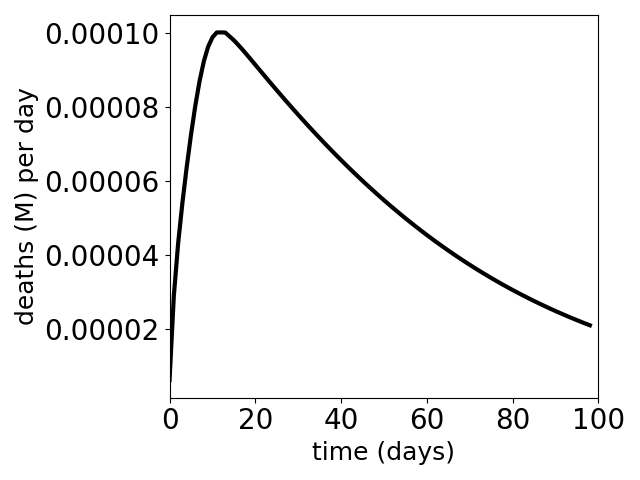}
		\caption{Number of deaths per day}
	\end{subfigure}
	\caption{Simulation results for the COVID-19 SEIR model with vaccination control and MTL specifications $\varphi_{\textrm{V}}^1$ (first row), $\varphi_{\textrm{V}}^2$ (second row) and $\varphi_{\textrm{V}}^3$ (third row).}  
	\label{fig_results}
\end{figure*}

We use the solver GEKKO \cite{beal2018gekko} to solve the optimization problems formulated in Section \ref{sec_control}. Fig. \ref{fig_results} and Table \ref{result_vaccination} show the simulation results for vaccination control of COVID-19 SEIR model with MTL specifications $\varphi_{\textrm{V}}^1$, $\varphi_{\textrm{V}}^2$ and $\varphi_{\textrm{V}}^3$, respectively. The results show that the MTL specifications $\varphi_{\textrm{V}}^1$, $\varphi_{\textrm{V}}^2$ and $\varphi_{\textrm{V}}^3$ are satisfied with the synthesized vaccination control inputs respectively. It can be seen that vaccination within the first 40 days after the outbreak can mitigate the spread of SARS-CoV-2 in the most efficient manner. The results also show that the control effort for satisfying $\varphi_{\textrm{V}}^1$ is less than that for satisfying $\varphi_{\textrm{V}}^2$, which is still less than that for satisfying $\varphi_{\textrm{V}}^3$. This is consistent with the fact that $\varphi_{\textrm{V}}^2$ implies $\varphi_{\textrm{V}}^1$, and $\varphi_{\textrm{V}}^3$ implies both $\varphi_{\textrm{V}}^1$ and $\varphi_{\textrm{V}}^2$. For all three specifications, the computations are completed within 4 seconds on a MacBook Laptop with 1.40-GHz Core i5 CPU and 16-GB RAM.

\subsection{COVID-19 SEIR Model with Shield Immunity Control}
\label{results_shield}
We use the same parameters of the COVID-19 SEIR model as shown in Table \ref{parameter_SEIR}. We also choose the same initial values of the states as $I[0]=1000$ (people), $E[0]=0.02$ million, $S[0]=9.979$ million, $R[0]=0$ and $D[0]=0$, with $S[0] + E[0] + I[0] + R[0]+D[0]=N_0=10$ million. We set $\chi_{\textrm{max}}=100$. The start time for the simulations in this subsection are February 23, 2020. We set the three MTL specifications $\varphi_{\textrm{S}}^1$, $\varphi_{\textrm{S}}^2$ and $\varphi_{\textrm{S}}^3$ (as shown in Table \ref{result_shield}) to be less stringent than the MTL specifications with the vaccination control, as shield immunity is generally less effective than vaccination. It can be shown that without any control strategies the three MTL specifications $\varphi_{\textrm{S}}^1$, $\varphi_{\textrm{S}}^2$ and $\varphi_{\textrm{S}}^3$ are all violated. Now we investigate the hypothetical scenario where the isolation measures are replaced by shield immunity control.

\begin{figure*}
	\centering
	\begin{subfigure}[b]{0.24\textwidth}
		\centering
		\includegraphics[width=\textwidth]{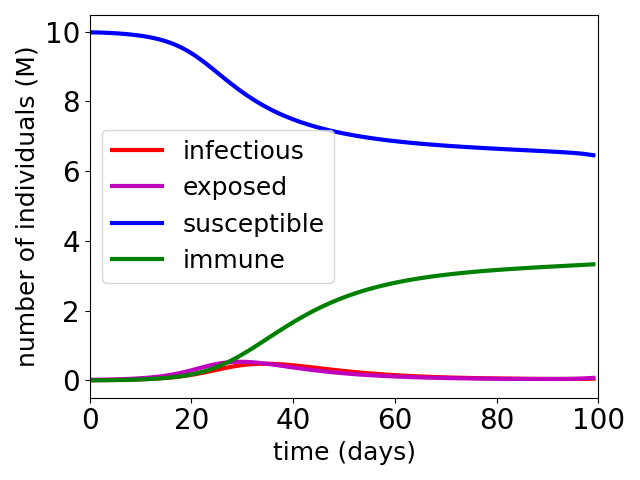}
% 		\caption{Number of individuals}
	\end{subfigure} 
	\hfill
	\begin{subfigure}[b]{0.24\textwidth}
		\centering
		\includegraphics[width=\textwidth]{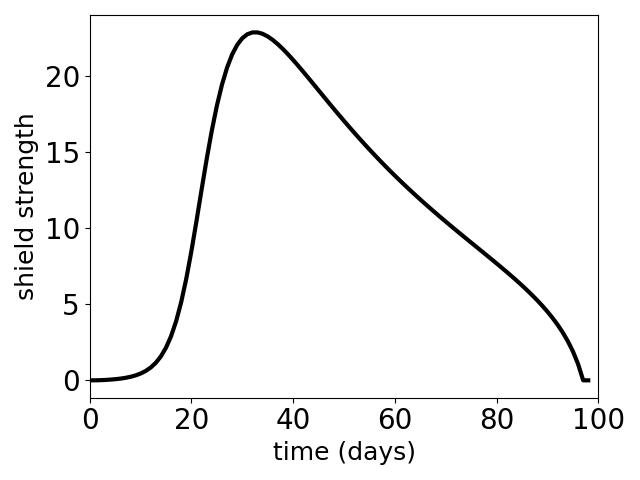}
%		\caption{Shield strength}
	\end{subfigure}
		\hfill
	\begin{subfigure}[b]{0.24\textwidth}
		\centering
		\includegraphics[width=\textwidth]{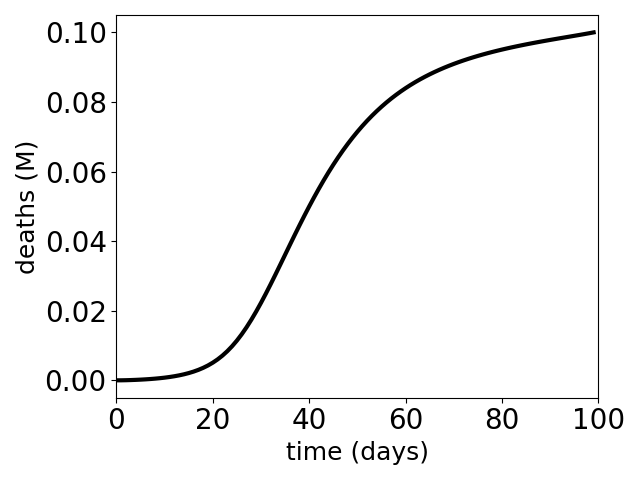}
%		\caption{Number of deaths}
	\end{subfigure}
		\hfill
	\begin{subfigure}[b]{0.24\textwidth}
		\centering
		\includegraphics[width=\textwidth]{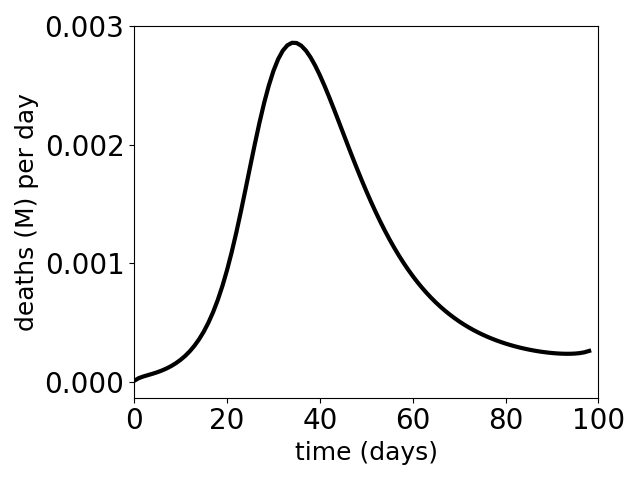}
%		\caption{Number of deaths per day}
	\end{subfigure}
\\
	\begin{subfigure}[b]{0.24\textwidth}
		\centering
		\includegraphics[width=\textwidth]{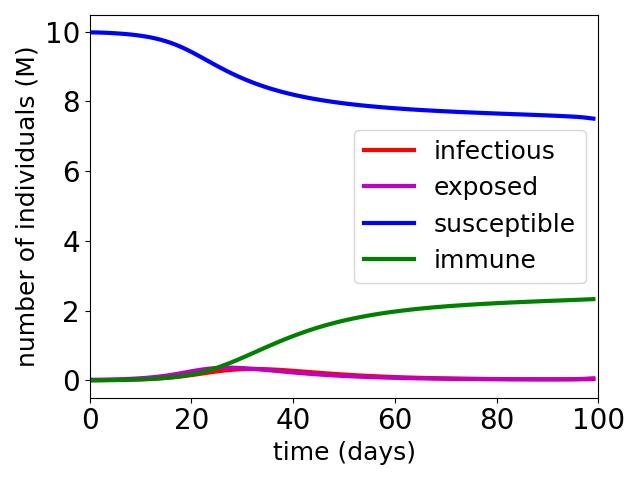}
%		\caption{Number of individuals}
	\end{subfigure} 
	\hfill
	\begin{subfigure}[b]{0.24\textwidth}
		\centering
		\includegraphics[width=\textwidth]{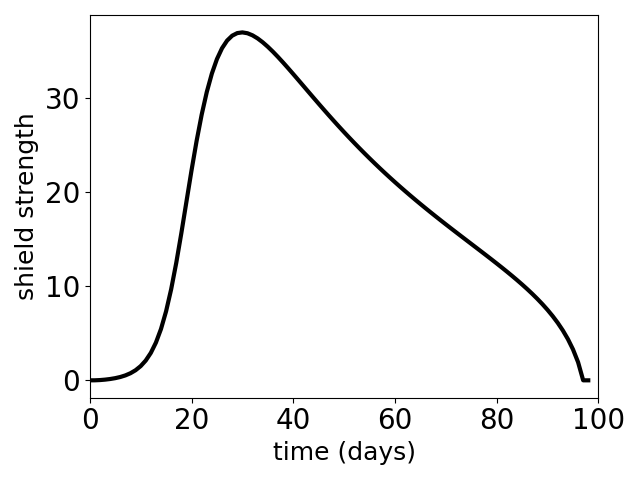}
%		\caption{Shield strength}
	\end{subfigure}
		\hfill
	\begin{subfigure}[b]{0.24\textwidth}
		\centering
		\includegraphics[width=\textwidth]{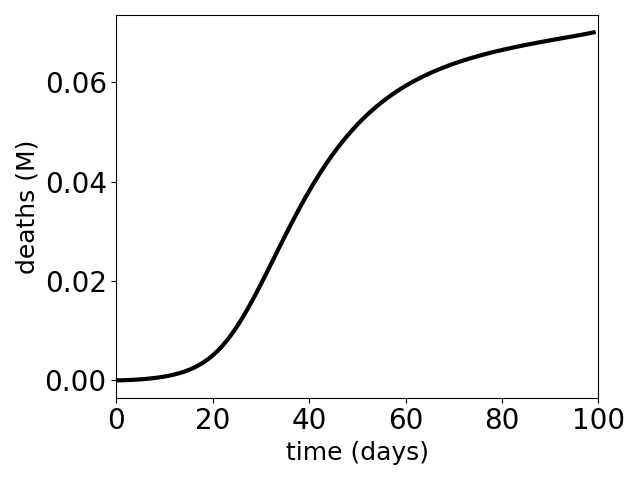}
%		\caption{Number of deaths}
	\end{subfigure}
		\hfill
	\begin{subfigure}[b]{0.24\textwidth}
		\centering
		\includegraphics[width=\textwidth]{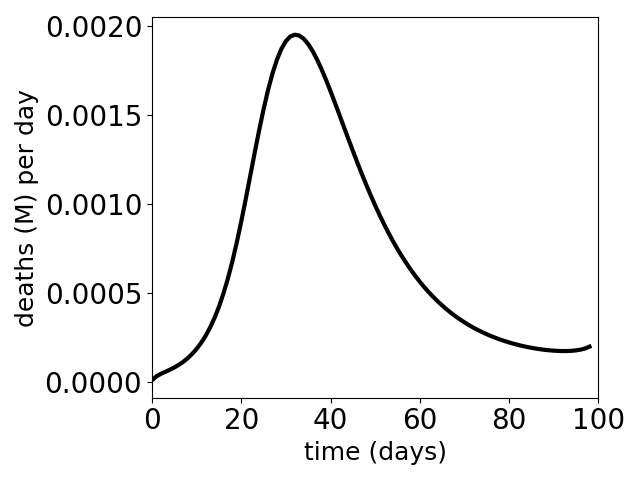}
%		\caption{Number of deaths per day}
	\end{subfigure}
\\
	\begin{subfigure}[b]{0.24\textwidth}
		\centering
		\includegraphics[width=\textwidth]{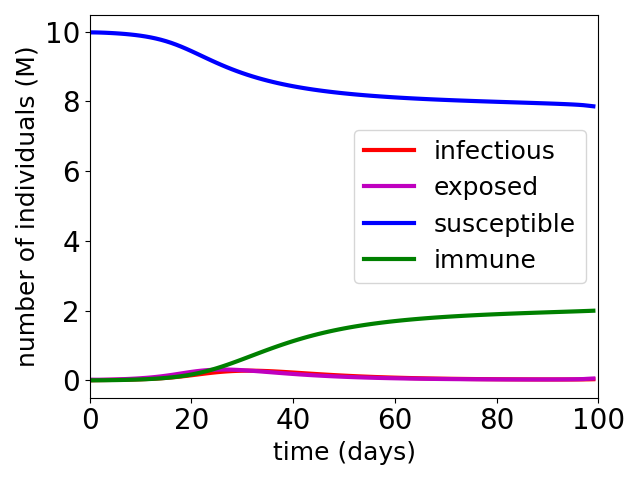}
		\caption{Number of individuals}
	\end{subfigure} 
	\hfill
	\begin{subfigure}[b]{0.24\textwidth}
		\centering
		\includegraphics[width=\textwidth]{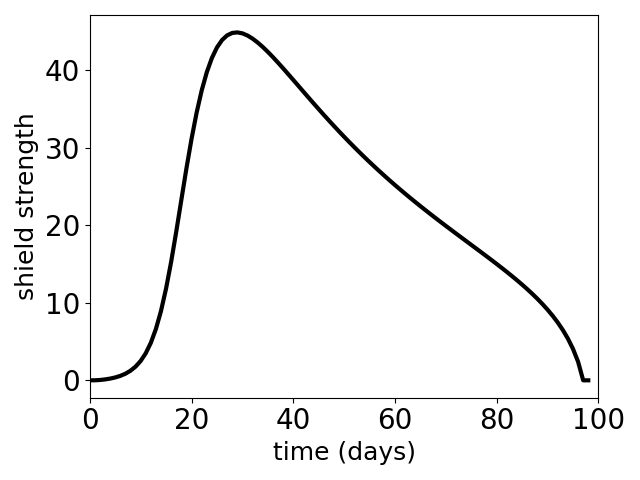}
		\caption{Shield strength}
	\end{subfigure}
		\hfill
	\begin{subfigure}[b]{0.24\textwidth}
		\centering
		\includegraphics[width=\textwidth]{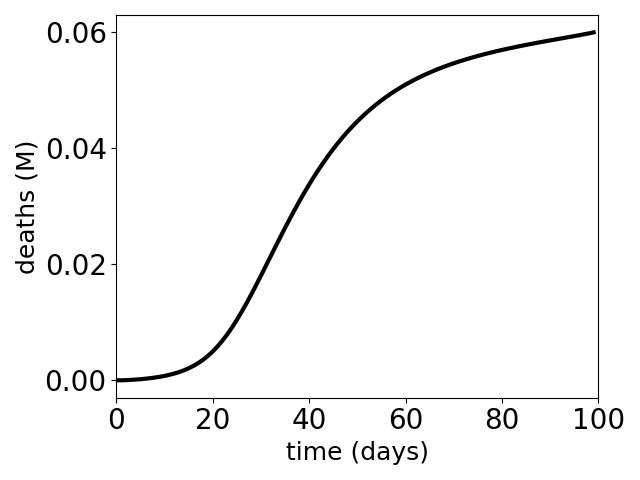}
		\caption{Number of deaths}
	\end{subfigure}
		\hfill
	\begin{subfigure}[b]{0.24\textwidth}
		\centering
		\includegraphics[width=\textwidth]{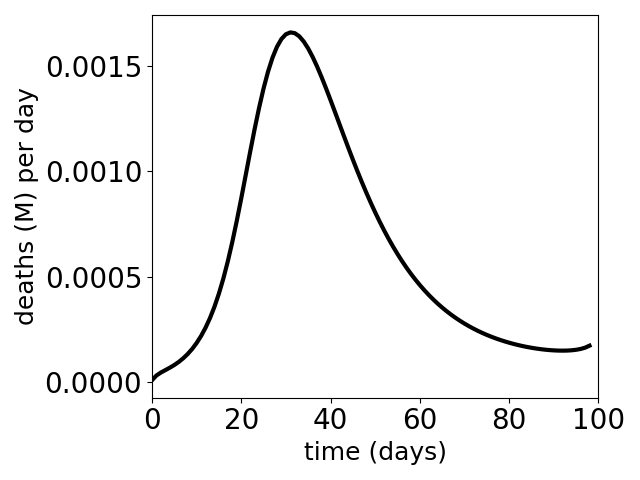}
		\caption{Number of deaths per day}
	\end{subfigure}
	\caption{Simulation results for the COVID-19 SEIR model with shield immunity control and MTL specifications $\varphi_{\textrm{S}}^1$ (first row), $\varphi_{\textrm{S}}^2$ (second row) and $\varphi_{\textrm{S}}^3$ (third row).}  
	\label{fig_results_shield}
\end{figure*}

Fig. \ref{fig_results_shield} and Table \ref{result_shield} show the simulation results for shield immunity control of the COVID-19 SEIR model with MTL specifications $\varphi_{\textrm{S}}^1$, $\varphi_{\textrm{S}}^2$ and $\varphi_{\textrm{S}}^3$, respectively. The results show that the MTL specifications $\varphi_{\textrm{S}}^1$, $\varphi_{\textrm{S}}^2$ and $\varphi_{\textrm{S}}^3$ are satisfied with the synthesized shield immunity control inputs respectively.
We observe that with the three MTL specifications, the synthesized shield immunity control inputs all increase to a peak after approximately 20 to 40 days and then gradually decrease. These observations indicate that shield immunity at early days of COVID-19 is more efficient than shield immunity at later days. The results also show that the control effort for satisfying $\varphi_{\textrm{S}}^1$ is less than that for satisfying $\varphi_{\textrm{S}}^2$, which is still less than that for satisfying $\varphi_{\textrm{S}}^3$. This is consistent with the fact that $\varphi_{\textrm{S}}^2$ implies $\varphi_{\textrm{S}}^1$, and $\varphi_{\textrm{S}}^3$ implies both $\varphi_{\textrm{S}}^1$ and $\varphi_{\textrm{S}}^2$. 

% We also tried control synthesis with respect to $\varphi_3$, but the solver returns ``infeasible'' after 250 iterations. 
	
    \begin{table}[]
		\centering
		\caption{MTL specifications and simulation results for shield immunity control (Section \ref{results_shield}).}
		\begin{tabular}{ll>{\raggedright\arraybackslash}p{20mm}}
		\toprule[2pt]    
			 ~~~~~MTL specification & \tabincell{c}{control\\ effort} & \tabincell{c}{computation\\ time}\\ \hline
		    \tabincell{c}{$\varphi_{\textrm{S}}^1=\Box_{[0,100]} (\Delta D\le0.003)$\\ $~~~~~~\wedge\Box_{[0,100]} (D\le0.1)$\\ $~~~\wedge\Diamond_{[40, 60]}(R\ge1)$}   & 16879.53 & ~~~2.112 s  \\ 
		     \tabincell{c}{$\varphi_{\textrm{S}}^2=\Box_{[0,100]} (\Delta D\le0.002)$\\ $~~~~~\wedge\Box_{[0,100]} (D\le0.07)$\\
		     $~~\wedge\Diamond_{[40, 60]}(R\ge1)$}  & 45595.10 & ~~~2.881 s \\ 
		     \tabincell{c}{$\varphi_{\textrm{S}}^3=\Box_{[0,100]} (\Delta D\le0.002)$\\ $~~~~~\wedge\Box_{[0,100]} (D\le0.06)$\\$\wedge\Diamond_{[40, 60]}(R\ge1)$} & 67786.88 & ~~~5.323 s \\ \bottomrule[2pt]       
		\end{tabular}
		\label{result_shield}  
	\end{table}

\subsection{COVID-19 SUQC Model with Quarantine Control}
\label{results_quarantine}
\begin{figure}
	\centering
	\begin{subfigure}[b]{0.24\textwidth}
		\centering
		\includegraphics[width=\textwidth]{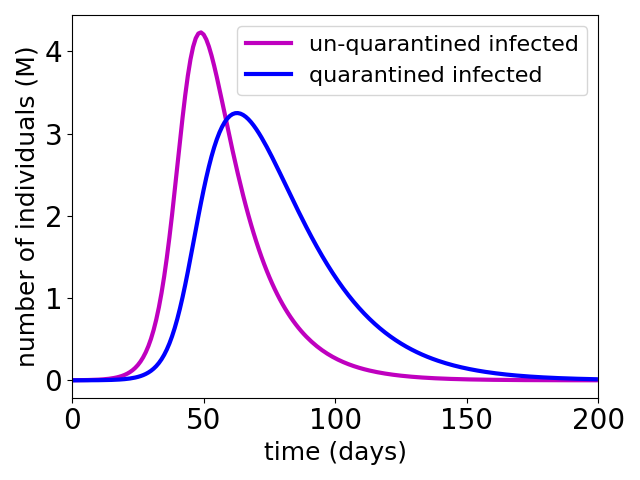}
		\caption{Number of un-quarantined and quarantined infected individuals}
	\end{subfigure} 
	\hfill
	\begin{subfigure}[b]{0.24\textwidth}
		\centering
		\includegraphics[width=\textwidth]{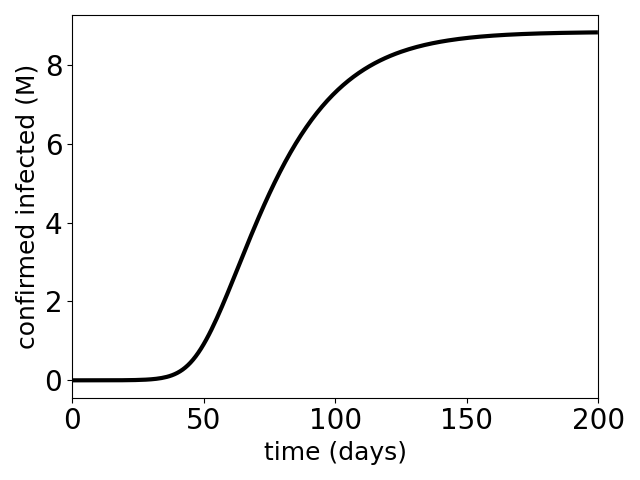}
		\caption{Number of confirmed infected individuals}
	\end{subfigure} 
	\hfill
	\begin{subfigure}[b]{0.24\textwidth}
		\centering
		\includegraphics[width=\textwidth]{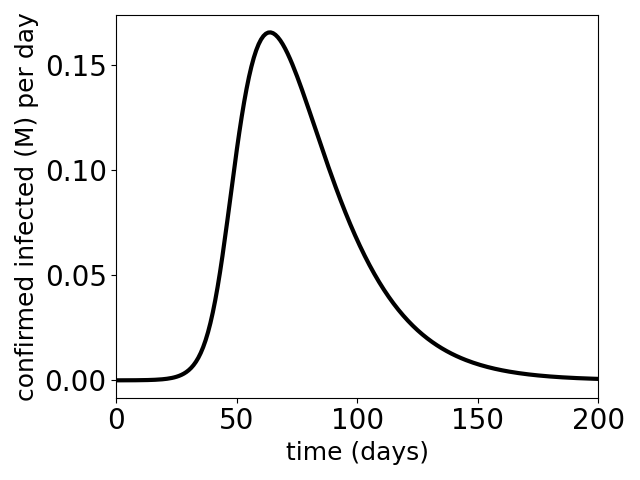}
		\caption{Number of confirmed infected individuals per day}
	\end{subfigure}
	\caption{Simulation results for the COVID-19 SUQC model estimated from data in Stage I of Wuhan, China.}  
	\label{fig_results_quarantine0}
\end{figure}

The parameters of the COVID-19 SUQC model are shown in Table \ref{parameter_SUQC}. They were estimated in \cite{Zhao2020} from the data in Wuhan, China. We choose the initial values of the states as $S[0]=8.9$ million, $U[0]=0.001$ million, $Q[0]=0$ and $C[0]=0$. We set $q_{\textrm{max}}=1$. We consider three MTL specifications as shown in Table \ref{result_quarantine}. For example, $\varphi_{\textrm{Q}}^1=\Box_{[0,200]} (\Delta C\le0.001)\wedge\Box_{[0,200]} (C\le0.1)$ means ``the confirmed infected population should never exceed 0.001 million (i.e., one thousand) per day and 0.1 million (i.e., 100 thousand) in total within the next 200 days''. The start time for the simulations in this subsection are January 20, 2020. Fig. \ref{fig_results_quarantine0} shows the simulation results for the COVID-19 SUQC model estimated from data in Stage I (January 20 to January 30, 2020) of Wuhan, China. 
It can be seen that the three MTL specifications $\varphi_{\textrm{Q}}^1$, $\varphi_{\textrm{Q}}^2$ and $\varphi_{\textrm{Q}}^3$ are all violated in such a situation (with quarantine rate being always 0.063). Now we investigate the scenario where the quarantine rate can be controlled to satisfy the MTL specifications.

\begin{figure*}
	\centering
	\begin{subfigure}[b]{0.24\textwidth}
		\centering
		\includegraphics[width=\textwidth]{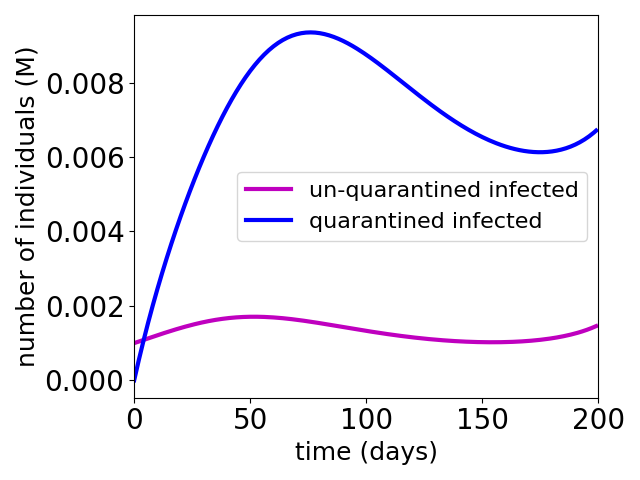}
%		\caption{Number of un-quarantined and quarantined infected individuals}
	\end{subfigure} 
	\hfill
	\begin{subfigure}[b]{0.24\textwidth}
		\centering
		\includegraphics[width=\textwidth]{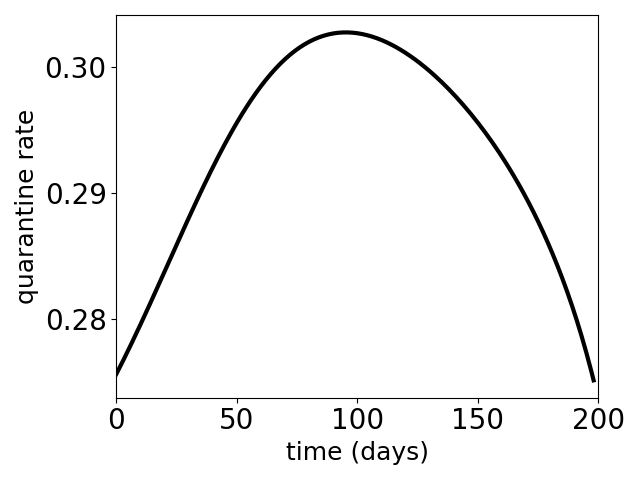}
%		\caption{Quarantine rate\vspace{3mm}}
	\end{subfigure}
	\hfill
	\begin{subfigure}[b]{0.24\textwidth}
		\centering
		\includegraphics[width=\textwidth]{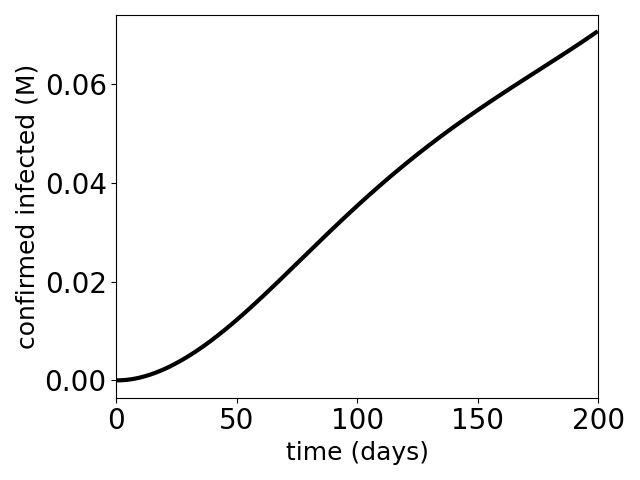}
%		\caption{Number of confirmed infected individuals}
	\end{subfigure} 
	\hfill
	\begin{subfigure}[b]{0.24\textwidth}
		\centering
		\includegraphics[width=\textwidth]{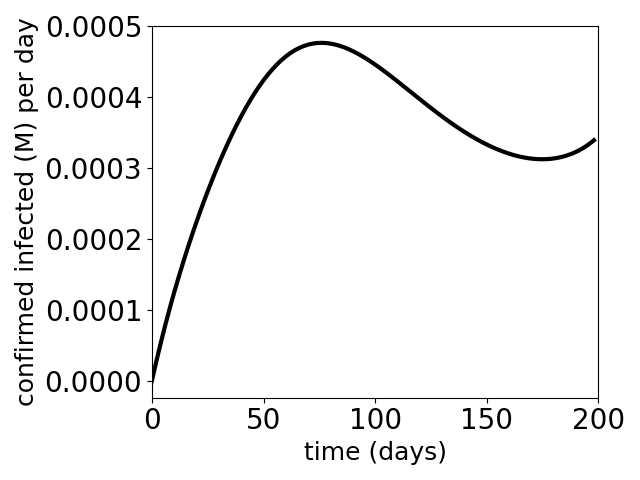}
%		\caption{Number of confirmed infected individuals per day}
	\end{subfigure}
\\
	\begin{subfigure}[b]{0.24\textwidth}
		\centering
		\includegraphics[width=\textwidth]{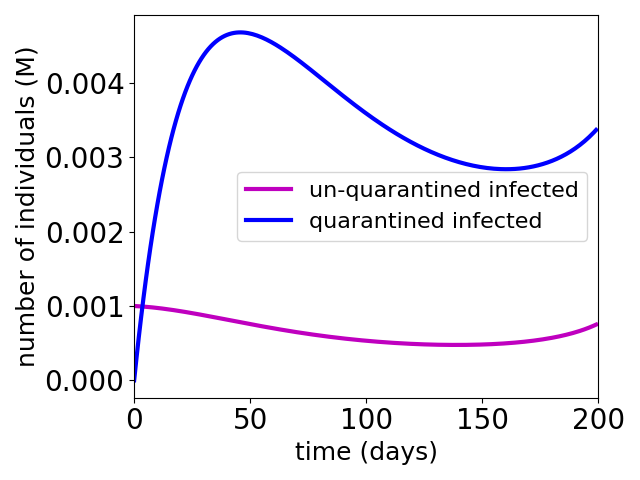}
%		\caption{Number of un-quarantined and quarantined infected individuals}
	\end{subfigure} 
	\hfill
	\begin{subfigure}[b]{0.24\textwidth}
		\centering
		\includegraphics[width=\textwidth]{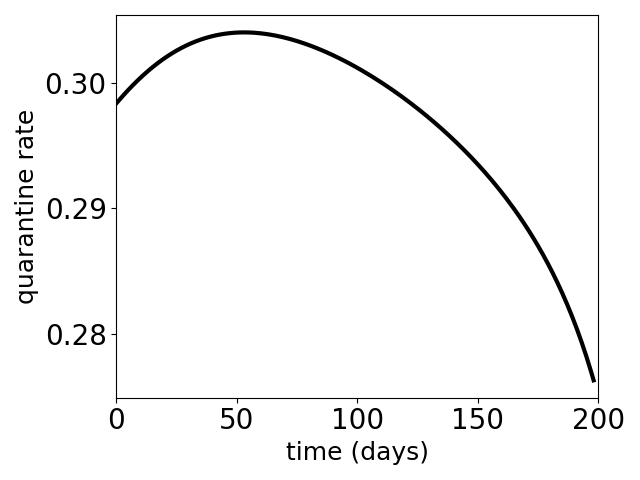}
%		\caption{Quarantine rate\vspace{3mm}}
	\end{subfigure}
	\hfill
	\begin{subfigure}[b]{0.24\textwidth}
		\centering
		\includegraphics[width=\textwidth]{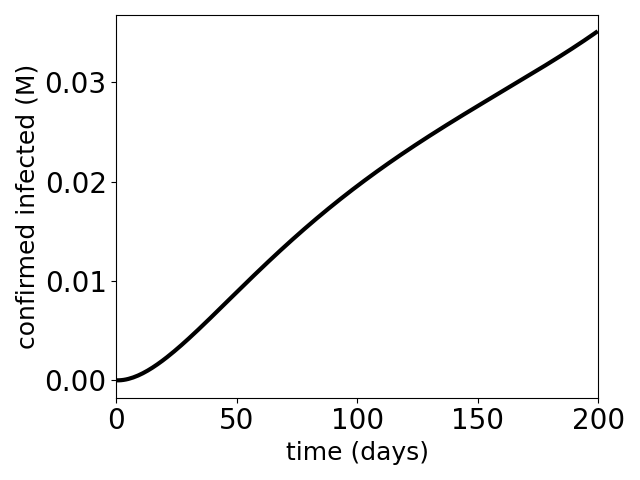}
%		\caption{Number of confirmed infected individuals}
	\end{subfigure} 
	\hfill
	\begin{subfigure}[b]{0.24\textwidth}
		\centering
		\includegraphics[width=\textwidth]{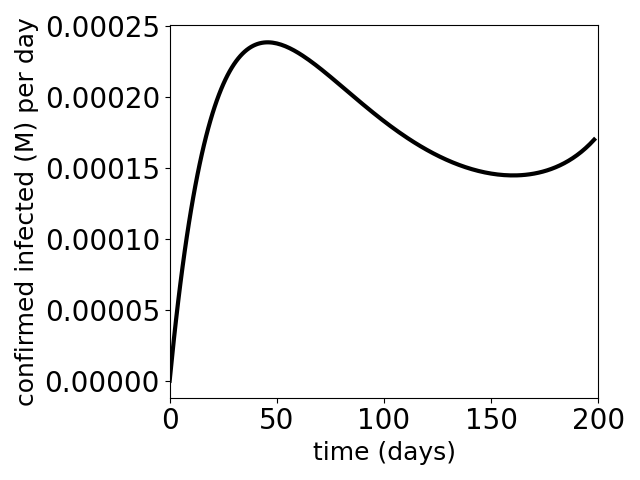}
%		\caption{Number of confirmed infected individuals per day}
	\end{subfigure}
\\
	\begin{subfigure}[b]{0.24\textwidth}
		\centering
		\includegraphics[width=\textwidth]{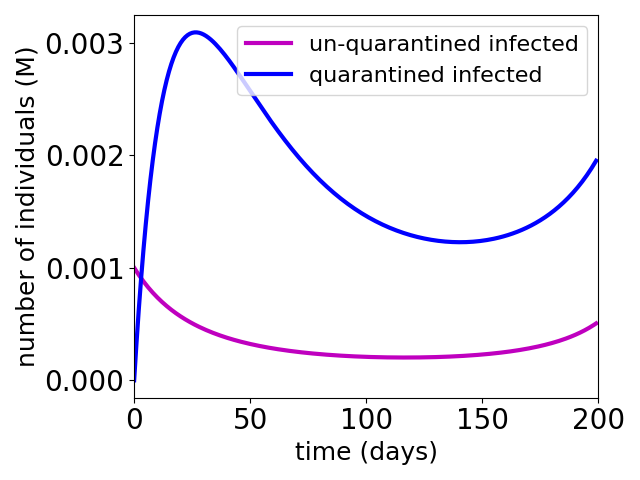}
		\caption{Number of un-quarantined and quarantined infected individuals}
	\end{subfigure} 
	\hfill
	\begin{subfigure}[b]{0.24\textwidth}
		\centering
		\includegraphics[width=\textwidth]{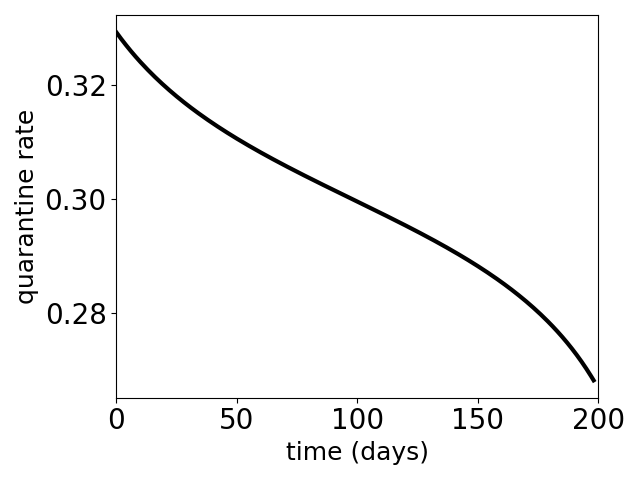}
		\caption{Quarantine rate\vspace{3mm}}
	\end{subfigure}
	\hfill
	\begin{subfigure}[b]{0.24\textwidth}
		\centering
		\includegraphics[width=\textwidth]{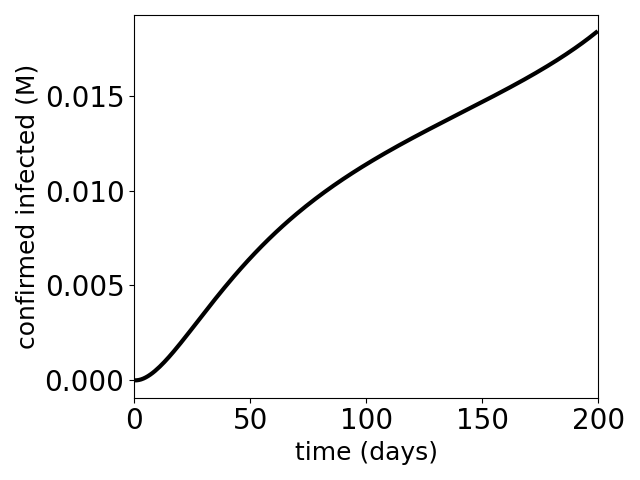}
		\caption{Number of confirmed infected individuals}
	\end{subfigure} 
	\hfill
	\begin{subfigure}[b]{0.24\textwidth}
		\centering
		\includegraphics[width=\textwidth]{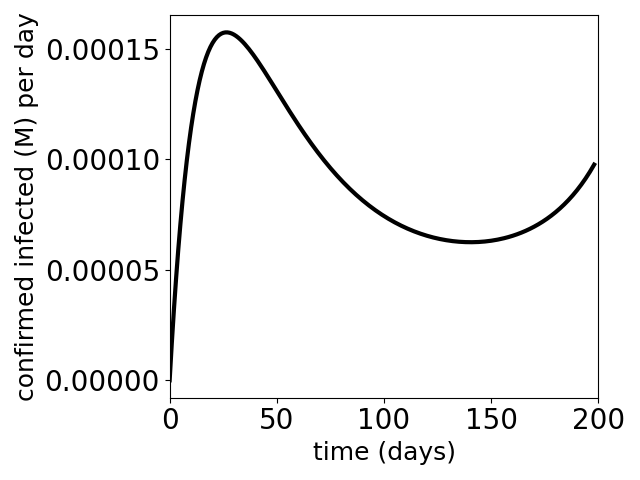}
		\caption{Number of confirmed infected individuals per day}
	\end{subfigure}
	\caption{Simulation results for the COVID-19 SUQC model with quarantine control and MTL specifications $\varphi_{\textrm{Q}}^1$ (first row), $\varphi_{\textrm{Q}}^2$ (second row) and $\varphi_{\textrm{Q}}^3$ (third row).}  
	\label{fig_results_quarantine}
\end{figure*}

Fig. \ref{fig_results_quarantine} and Table \ref{result_quarantine} show the simulation results for quarantine control of the COVID-19 SUQC model with MTL specifications $\varphi_{\textrm{Q}}^1$, $\varphi_{\textrm{Q}}^2$ and $\varphi_{\textrm{Q}}^3$, respectively. The results show that the MTL specifications $\varphi_{\textrm{Q}}^1$, $\varphi_{\textrm{Q}}^2$ and $\varphi_{\textrm{Q}}^3$ are satisfied with the synthesized quarantine control inputs respectively. The results also
show that the control effort for satisfying $\varphi_{\textrm{Q}}^1$ is less than that for satisfying $\varphi_{\textrm{Q}}^2$, which is still less than that for satisfying $\varphi_{\textrm{Q}}^3$. This is consistent with the fact that $\varphi_{\textrm{Q}}^2$ implies $\varphi_{\textrm{Q}}^1$, and $\varphi_{\textrm{Q}}^3$ implies both $\varphi_{\textrm{Q}}^1$ and $\varphi_{\textrm{Q}}^2$. 
We observe that with $\varphi_{\textrm{Q}}^1$, the synthesized quarantine control inputs first increase to a peak at approximately 90 days and then gradually decrease; with $\varphi_{\textrm{Q}}^2$, the synthesized quarantine control inputs first increase to a peak at approximately 50 days and then gradually decrease; and with $\varphi_{\textrm{Q}}^3$, the synthesized quarantine control inputs are at a peak from the beginning and gradually decrease. These observations indicate that quarantine in the early days of COVID-19 can reduce the number of confirmed infected cases more efficiently than quarantine in the later days, and more stringent control specifications require stronger quarantine measures to be implemented.

	\begin{table}[]
		\centering
		\caption{Parameters of the COVID-19 SUQC model estimated from data in Stage I (January 20 to January 30, 2020) of Wuhan, China \cite{Zhao2020}.}
		\label{parameter_SUQC}
		\begin{tabular}{llll}
			\toprule[2pt]    
			parameter    & value  & parameter    & value \\ \hline
		    ~~~~$\beta_0$        & 0.2967  & ~~~~$\gamma_2$        & 0.05\\ 
		    ~~~~$\hat{N}_0$        & 8.9 million & ~~~~$\sigma$        & 0.001\\
		    ~~~~$T_{\textrm{s}}$  & 1 day \\
         \bottomrule[2pt]   
		\end{tabular}
	
	\end{table}  

    \begin{table}[]
		\centering
		\caption{MTL specifications and simulation results for quarantine control (Section \ref{results_quarantine}).}
		\begin{tabular}{ll>{\raggedright\arraybackslash}p{20mm}}
			\toprule[2pt]    
			 ~~~~~MTL specification & \tabincell{c}{control\\ effort} & \tabincell{c}{computation\\ time}\\ \hline
		    \tabincell{c}{$\varphi_{\textrm{Q}}^1=\Box_{[0,200]} (\Delta C\le0.001)$\\ $~~~~~~\wedge\Box_{[0,200]} (C\le0.1)$}   & ~19.502 & ~~2.296 s  \\ 
		     \tabincell{c}{$\varphi_{\textrm{Q}}^2=\Box_{[0,200]} (\Delta C\le0.0005)$\\ $~~~~~\wedge\Box_{[0,200]} (C\le0.05)$}  & ~20.023 & ~~2.598 s \\ 
		     \tabincell{c}{$\varphi_{\textrm{Q}}^3=\Box_{[0,200]} (\Delta C\le0.0005)$\\ $~~~~~\wedge\Box_{[0,200]} (C\le0.03)$} & ~20.3 & ~~4.578 s \\ \bottomrule[2pt]       
		\end{tabular}
		\label{result_quarantine}  
	\end{table}

\section{Conclusion}
In this paper, we proposed a systematic control synthesis approach for mitigating the COVID-19 epidemic based on three control models with vaccination, shield immunity and quarantine, respectively. We used metric temporal logic (MTL) formulas to formally specify the required performance of the control strategies. The proposed approach can synthesize control inputs that lead to satisfaction of the MTL specifications. 

The work in this paper opens the door to the formal synthesis of control strategies based on epidemic models. We list several future directions as follows. First, we will investigate the effects of model uncertainties and parameter uncertainties in the control synthesis, and explore robust control synthesis methods in the presence of such uncertainties. Second, we will extend this work to online control synthesis so that control inputs can be generated in real-time based on the latest information (e.g., using online parameter identification and receding horizon control). Finally, as we investigated the three control strategies separately in this paper, we will study the benefits and costs of joint control of different control strategies (vaccination, shield immunity and quarantine) so that the specifications can be satisfied with coordinated efforts.

\bibliographystyle{IEEEtran}
\bibliography{ref}

\begin{IEEEbiography}[{\includegraphics[width=1in,height=1.25in,clip,keepaspectratio]{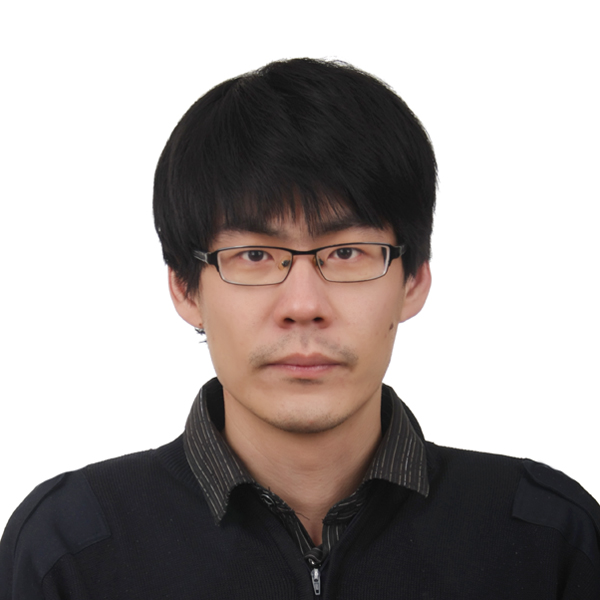}}]{Zhe Xu}
received the B.S. and M.S. degrees in Electrical Engineering from Tianjin University, Tianjin, China, in 2011 and 2014, respectively. He received the Ph.D. degree in Electrical Engineering at Rensselaer Polytechnic Institute, Troy, NY, in 2018. He is currently a postdoctoral researcher in the Oden Institute for Computational Engineering and Sciences at the University of Texas at Austin, Austin, TX. He will join the School for Engineering of Matter, Transport and Energy at Arizona State University as an assistant professor in January 2021. His research interests include formal methods, autonomous systems, biological systems, control systems and reinforcement learning. 
\end{IEEEbiography}

\begin{IEEEbiography}[{\includegraphics[width=1in,height=1.25in,clip,keepaspectratio]{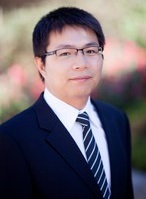}}]{Bo Wu}
received his B.E. degree from Harbin Institute of Technology, China, in 2008, an M.S. degree from Lund University, Sweden, in 2011 and Ph.D. degree from the University of Notre Dame, USA, in 2018, all in electrical engineering. He is currently a postdoctoral researcher at the Oden Institute for Computational Engineering and Sciences at the University of Texas at Austin. His research interest is to apply formal methods, learning, and control in autonomous systems, such as robotic systems, communication systems, and human-in-the-loop systems, to provide privacy, security, and performance guarantees.
\end{IEEEbiography}

\begin{IEEEbiography}[{\includegraphics[width=1in,height=1.25in,clip,keepaspectratio]{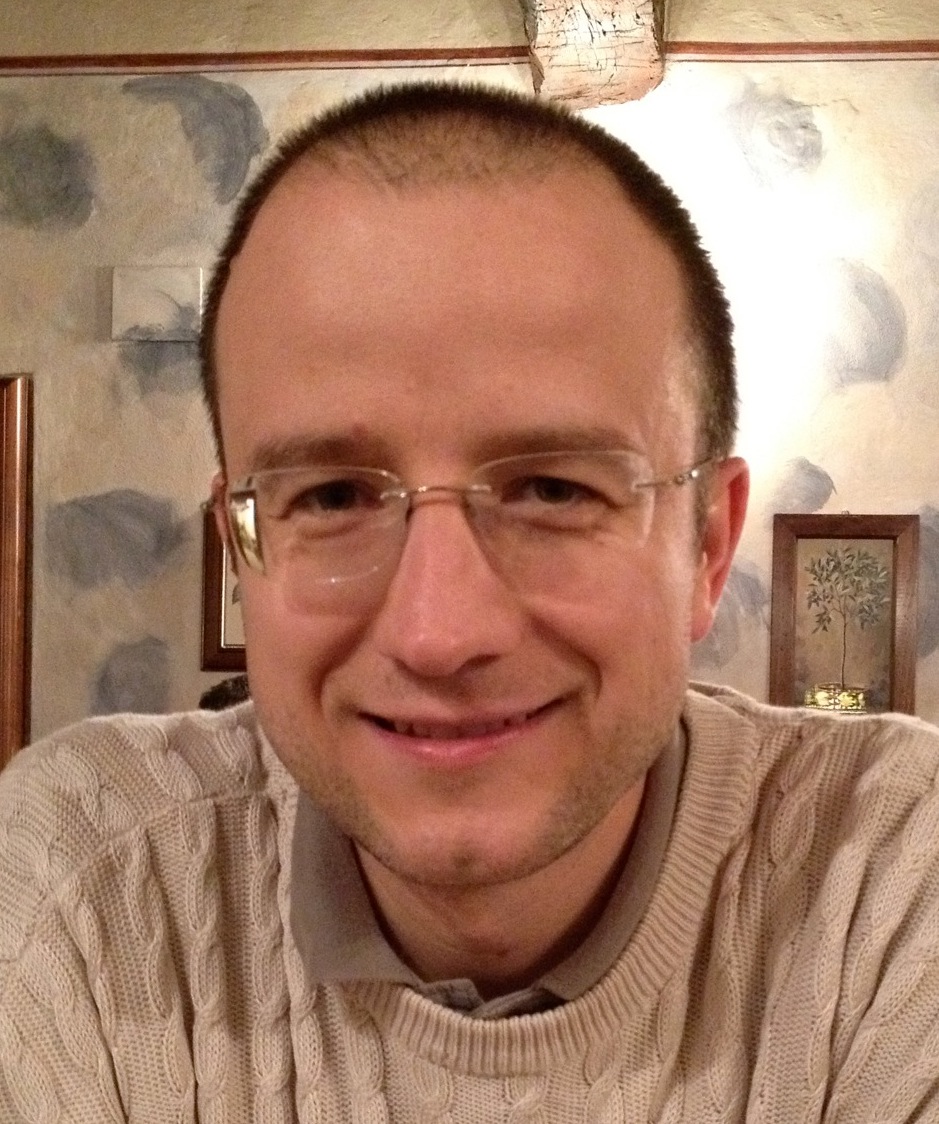}}]{Ufuk Topcu}
Ufuk Topcu joined the Department of Aerospace Engineering at the University of Texas at Austin as an assistant professor in Fall 2015. He received his Ph.D. degree from the University of California at Berkeley in 2008. He held research positions at the University of Pennsylvania and California Institute of Technology. His research focuses on the theoretical, algorithmic and computational aspects of design and verification of autonomous systems through novel connections between formal methods, learning theory and controls. 
\end{IEEEbiography}

\end{document}